\documentclass{article}

\usepackage{arxiv}

\usepackage[utf8]{inputenc}
\usepackage[T1]{fontenc}
\usepackage{hyperref}
\usepackage{url}
\usepackage{booktabs}
\usepackage{amsmath}
\usepackage{amsfonts}
\usepackage{nicefrac}
\usepackage{microtype}
\usepackage[capitalize]{cleveref}
\usepackage{lipsum}
\usepackage{graphicx}
\usepackage[style=ieee,sorting=none,isbn=false,url=false]{biblatex}
\usepackage{doi}
\usepackage{orcidlink}
\usepackage[acronyms,nonumberlist]{glossaries-extra}
\usepackage{mathtools}
\usepackage{physics}
\usepackage{comment}
\usepackage{titlesec}
\usepackage{multirow}
\usepackage{siunitx}

\setabbreviationstyle[acronym]{long-short}

\newacronym{ads}{ADS}{automatic domain splitting}

\newacronym{bvp}{BVP}{boundary value problem}

\newacronym{ci}{CI}{confidence interval}
\newacronym{cdf}{CDF}{cumulative distribution function}
\newacronym{cl}{CL}{confidence level}
\newacronym{cnes}{CNES}{Centre National d'\'Etudes Spatiales}

\newacronym{da}{DA}{differential algebra}
\newacronym{dim}{DIM}{Doppler integration method}
\newacronym[\glslongpluralkey={degrees of freedom}]{dof}{DoF}{degree of freedom}

\newacronym{fp}{FP}{floating point}

\newacronym{ic}{IC}{initial condition}
\newacronym{iod}{IOD}{initial orbit determination}

\newacronym{leo}{LEO}{Low Earth Orbit}
\newacronym{los}{LOS}{line of sight}
\newacronym{ls}{LS}{least squares}

\newacronym{ols}{OLS}{ordinary least squares}

\newacronym{pace}{PACE}{Polynomial Algebra Computational Engine}

\newacronym{so}{SO}{space object}

\newacronym{tle}{TLE}{Two Line Elements}

\newacronym{wls}{WLS}{weighted least squares}
\newcommand{\glsentryfullinv}[1]{\glsentryshort{#1} (\glsentrylong{#1})}

\definecolor{AddedColor}{HTML}{008ae6}
\definecolor{DeletedColor}{HTML}{ff0000}

\newcommand{\deleted}[1]{}

\title{Perturbed Initial Orbit Determination}

\date{}

\author{Alberto Fossà\thanks{PhD Candidate, Department of Aerospace Vehicles Design and Control, 10 Avenue Edouard Belin.}\enspace\orcidlink{0000-0002-0756-4998} \\
	Institut Supérieur de l'Aéronautique et de l'Espace\\
	31055 Toulouse, France\\
	\href{mailto:alberto.fossa@isae-supaero.fr}{\texttt{alberto.fossa@isae-supaero.fr}} \\
	\And
	Matteo Losacco\thanks{Postdoctoral Researcher, Department of Aerospace Vehicles Design and Control, 10 Avenue Edouard Belin.}\enspace\orcidlink{0000-0002-5126-1724} \\
	Institut Supérieur de l'Aéronautique et de l'Espace\\
	31055 Toulouse, France\\
	\href{mailto:matteo-los@hotmail.it}{\texttt{matteo-los@hotmail.it}} \\
        \And
        Roberto Armellin\thanks{Full Professor, Te P\=unaha \=Atea - Space Institute, 20 Symonds Street.}\enspace\orcidlink{0000-0002-3516-6428}\\
        University of Auckland\\
        1010 Auckland, New Zealand\\
        \href{mailto:roberto.armellin@auckland.ac.nz}{\texttt{roberto.armellin@auckland.ac.nz}}
}

\renewcommand{\headeright}{}
\renewcommand{\undertitle}{}
\renewcommand{\shorttitle}{Perturbed Initial Orbit Determination}

\hypersetup{
pdftitle={Perturbed Initial Orbit Determination},
pdfsubject={},
pdfauthor={Alberto Fossà, Matteo Losacco, Roberto Armellin},
pdfkeywords={initial orbit determination, range radar, Doppler radar, optical telescope, differential algebra},
colorlinks=true,
allcolors=cyan
}

\begin{document}
\maketitle

\begin{abstract}
An algorithm for robust initial orbit determination (IOD) under perturbed orbital dynamics is presented. By leveraging map inversion techniques defined in the algebra of Taylor polynomials, this tool returns a highly accurate solution to the IOD problem and estimates a range centered on the aforementioned solution in which the true orbit should lie. To meet the specified accuracy requirements, automatic domain splitting is used to wrap the IOD routines and ensure that the local truncation error, introduced by a polynomial representation of the state estimate, remains below a predefined threshold. The algorithm is presented for three types of ground-based sensors, namely range radars, Doppler-only radars, and optical telescopes, by considering their different constraints in terms of available measurements and sensor noise. Finally, the improvement in performance with respect to a Keplerian-based IOD solution is demonstrated using large-scale numerical simulations over a subset of tracked objects in low Earth orbit.
\end{abstract}


\section{Introduction}
\label{sec:introduction}
An accurate characterization of the environment around the Earth is of paramount importance for all the operations required to ensure the safety of in-orbit missions such as observation scheduling, collision risk assessment, and re-entry predictions. This characterization typically consists in estimating and updating the state and attitude of any active and inactive orbital object, and it can be performed only if the lasts are observed with sufficient accuracy and frequency. As a result, ground- and space-based optical, radar, and laser sensors are jointly used to monitor the near-Earth environment.

Nevertheless, the discrepancies between regularly tracked objects and those predicted by mathematical models are significant. According to the 2022 ESA space debris environment report~\cite{ESASDO2021}, approximately 36,500 space debris with size greater than 10~cm are estimated to exist. This number soars when considering smaller dimensions, with 1 million fragments between 1~cm and 10~cm, and 130 million pieces between 1~mm and 1~cm. Only around 33,640 objects are regularly tracked by space surveillance networks and maintained in their catalogs. That is, existing catalogs currently cover just the larger-size portion of the orbiting population. Although the size and properties of the remaining objects prevent existing operational sensors from detecting or observing them with sufficient accuracy, the relentless launch activity and in-orbit generation events, coupled with the constant technological improvements in modern sensors, always offer the possibility of identifying and potentially characterizing uncatalogued objects.

Whenever an uncatalogued object is detected, an attempt is made to characterize its orbital state starting from the set of available measurements. If the so-called \glsentryfull{iod} process is successful, the object is associated with a state estimate, but not yet cataloged. New observations on subsequent passages are required to refine this estimate before a new entry is added to the catalog. The methods to perform \glsentryshort{iod} can generally be divided according to the type of available measurements and the sensor that generates them. There are three main categories of sensors: optical telescopes, range radars, and Doppler-only radars. Optical telescopes provide very accurate measurements of the angular position of a transiting object at each instant of observation. The resulting so-called ``angle-only'' methods exploit this information to estimate the state of transiting objects at one observation epoch. Classical examples of angle-only \glsentryshort{iod} methods include Laplace's method~\cite{Laplace1780}, Gauss' method~\cite{Gauss1809}, the double R iteration method~\cite{Escobal1965}, Baker-Jacobi's method~\cite{BakerJr1977}, Gooding's method~\cite{Gooding1996}, and Karimi and Mortari's methods~\cite{Karimi2010}. Range radars are characterized by less precise angular measurements but couple this information with the measurement of the object distance from the sensor, called slant range. Examples of classical ``angle-range'' \glsentryshort{iod} methods are Lambert's method~\cite{Izzo2015}, Gibbs' method~\cite{Gibbs1889}, and Herrick-Gibbs' method~\cite{Herrick1971}. Instead, Doppler-only radars combine angular information with the time derivative of the slant range. Examples of tailored \glsentryshort{iod} methods include the \glsentryfull{dim}~\cite{Yanez2017} and the hodograph method~\cite{Christian2021}.

One limitation of these algorithms is the lack of information regarding the uncertainty on the provided state estimate, which is however of great importance for quantifying the accuracy of the obtained solution and performing data association to update the available catalogs. Recent \glsentryshort{iod} methods leveraged \glsentryfull{da} techniques to overcome this problem by computing a polynomial expansion of the state estimate with respect to the uncertainties in the available measurements. Moreover, formulating the \glsentryshort{iod} problem in the \glsentryshort{da} framework eliminates the need for an iterative procedure to solve for the object state, since polynomial map inversion is exploited to solve the implicit equations that arise from the problem formulation. Given the local validity of truncated Taylor series, the \glsentryshort{da}-based \glsentryshort{iod} methods must be coupled with the \glsentryfull{ads} algorithm~\cite{Wittig2015a} to control the truncation error introduced by a fixed expansion order. The final output is thus a manifold (i.e., a set) of polynomials, each of them defined in a specific subdomain, whose union describes the uncertainty on the \glsentryshort{iod} estimate with the required accuracy. These methods, named DAIOD, were developed for all three types of ground-based sensors, namely optical telescopes~\cite{Pirovano2020e}, range radars~\cite{Armellin2018}, and Doppler-only radars~\cite{Losacco2023}, and constitute compelling alternatives to the more widespread \glsentryshort{iod} techniques cited above. Their advantages over classical methods were demonstrated in~\cite{Pirovano2020e} by comparing the DAIOD and Gooding's methods for angle-only \glsentryshort{iod}.

To obtain a solution, all previously cited methods resort to a simplified dynamical model, usually unperturbed Keplerian motion, which potentially reduces their range of applicability. Although generally valid for short-arc observations, this assumption no longer holds for longer arcs, and its introduction can affect the accuracy of both the orbital solution and its estimated uncertainty, with possible undesirable effects at a later stage during the data association process.

Based on these considerations, this study introduces three new \glsentryshort{iod} methods for angle-only, angle-range, and angle-range rate observations. The algorithms build on the aforementioned \glsentryshort{da}-based \glsentryshort{iod} methods and extend them to generic perturbed dynamics. 
The remainder of this paper is organized as follows. \Cref{sec:math_background} introduces the required mathematical background, namely \glsentryshort{da}, \glsentryshort{ads} and measurement regression. \Cref{sec:j2_pert_iod} illustrates the three methods for range radars (\cref{sec:rr}), Doppler-only radars (\cref{sec:dr}), and optical telescopes (\cref{sec:o}). Finally, \cref{sec:num_simu} presents the results of the numerical simulations and compares the proposed perturbed DAIOD methods with their Keplerian counterparts.

\section{Mathematical background}\label{sec:math_background}

This section introduces the main mathematical tools on which this study is based. They include \glsentryshort{da} with a focus on map inversion techniques, \glsentryshort{ads} for controlling the truncation error, and a brief discussion of the dynamical model used to perform \glsentrylong{iod}.

\subsection{\texorpdfstring{\Glsentrylong{da}}{Differential algebra}}\label{sec:da}

\Glsentrylong{da} is a computing technique that stems from the idea that it is possible to extract more information from a function $f$ than its mere value $f(x)$ at a point $x$. Given any function $\vb*{f}:\mathbb{R}^n\to\mathbb{R}^m$ that is $\mathcal{C}^{k+1}$ in the domain of interest $\mathcal{D}=[-1,1]^n$, the algebra of \glsentrylong{fp} numbers is replaced by a new algebra of Taylor polynomials to compute the $k^{th}$ order expansion of $\vb*{f}$~\cite{Berz1999}. The notation used in this study is as follows
\begin{equation}
    \vb*{f} \approx \left[\vb*{f}\right] = \mathcal{T}_{\vb*{f}}(\delta\vb*{x})
\end{equation}
where $\delta\vb*{x}=\{\delta x_1,\ldots,\delta x_n\}^T$ are the $n$ independent \glsentryshort{da} variables. When working with physical quantities $\vb*{x}$, it is convenient to introduce scaling factors $\vb*{\beta}$ such that $\delta\vb*{x}\in\mathcal{D}$. The domain of $\vb*{x}$ is then represented in the \glsentryshort{da} framework as
\begin{equation}
    [\vb*{x}] = \bar{\vb*{x}} + \vb*{\beta}\odot\delta\vb*{x}
\end{equation}
where $\bar{\vb*{x}}$ is the nominal value of $\vb*{x}$, $\odot$ denotes the Hadamard product, $\vb*{\beta}\in\mathbb{R}^n_{\geq 0}$ and $\delta\vb*{x}\in\mathcal{D}$. The parameters $\vb*{\beta}$ may have different physical meanings. For instance, if $\vb*{x}$ is Gaussian distributed with mean $\bar{\vb*{x}}$ and diagonal covariance $\vb*{\Sigma}=\text{diag}(\sigma_i)$, $\beta_i$ is commonly set equal to $3\sigma_i$ such that $[\vb*{x}]$ represents a domain spanning three standard deviations around its mean $\bar{\vb*{x}}$. In the specific case of \glsentryshort{iod}, values for $\vb*{\beta}$ are given by either \cref{eq:da_init_ci} or \cref{eq:da_init_3s}.

The four arithmetic operations, elementary functions (e.g., exponential, logarithm and trigonometric functions), derivation, integration, map composition and inversion are all well defined in \glsentryshort{da}. These basic operations can then be combined to derive powerful algorithms for the solution of implicit equations, the computation of the flow of the dynamics in terms of their \glsentrylongpl{ic}~\cite{Valli2013}, and the solution of \glsentrylongpl{bvp}~\cite{Armellin2018h}.

\subsubsection{Map inversion}

Suppose that the $k^{th}$-order Taylor expansion of $\vb*{y}$ in terms of $\delta\vb*{x}$ is known as $[\vb*{y}]=\mathcal{T}_{\vb*{y}}(\delta\vb*{x})$, but an explicit expression for its inverse is needed instead. This problem can be efficiently solved in the \glsentryshort{da} framework using a simple fixed-point iteration scheme~\cite{Berz1999}. The polynomial $\mathcal{T}_{\vb*{y}}(\delta\vb*{x})$ is firstly split into its constant and nonconstant parts as
\begin{equation}
    \mathcal{T}_{\vb*{y}}(\delta\vb*{x}) = \bar{\vb*{y}}+\mathcal{T}_{\delta\vb*{y}}(\delta\vb*{x})
\end{equation}
Then, the map $\delta\vb*{y}=\mathcal{T}_{\delta\vb*{y}}(\delta\vb*{x})$ is inverted as follows: its polynomial expansion is further separated into its linear and nonlinear parts as
\begin{equation}
    \mathcal{T}_{\delta\vb*{y}} = M_{\delta\vb*{y}}+\mathcal{N}_{\delta\vb*{y}}
\end{equation}
where the dependency on $\delta\vb*{x}$ is omitted for clarity. It is then observed that
\begin{equation}
    \mathcal{T}_{\delta\vb*{y}}\circ\mathcal{T}^{-1}_{\delta\vb*{y}} = \mathcal{I}
\end{equation}
with $\mathcal{I}$ the identity map, $\mathcal{T}^{-1}_{\delta\vb*{y}}$ the inverse map, and $\circ$ denotes map composition. A fixed-point scheme can then be setup as
\begin{equation}
    \mathcal{T}^{-1}_{\delta\vb*{y}} = M^{-1}_{\delta\vb*{y}}\circ\left(\mathcal{I}-\mathcal{N}_{\delta\vb*{y}}\circ\mathcal{T}^{-1}_{\delta\vb*{y}}\right)
    \label{eq:map_inv_fix_pt}
\end{equation}
the inverse map is guaranteed to exist if $M_{\delta\vb*{y}}$ is invertible, in which case \cref{eq:map_inv_fix_pt} converges in exactly $k$ steps, where $k$ is the order of expansion. The Taylor expansion of $\delta\vb*{x}$ is finally obtained as
\begin{equation}
    \delta\vb*{x}=\mathcal{T}_{\delta\vb*{x}}(\delta\vb*{y})=\mathcal{T}^{-1}_{\delta\vb*{y}}(\delta\vb*{y})
\end{equation}
This algorithm is extensively used in \cref{sec:j2_pert_iod} to compute a correction to the \glsentryshort{iod} solution that guarantees a continuous trajectory in the perturbed dynamical model of choice.

\subsection{\texorpdfstring{\Glsentrylong{ads}}{Automatic domain splitting}}\label{sec:ads}

Taylor polynomials are only local approximations of the function $\vb*{f}$ around its expansion point $\bar{\vb*{x}}$, and the accuracy of the \glsentryshort{da} map $\mathcal{T}_{\vb*{y}}(\delta\vb*{x})$ decreases when moving farther from $\bar{\vb*{x}}$. Given the domain of interest for $\delta\vb*{x}$, typically $\delta\vb*{x}\in[-1,1]^n$, the objective is to maintain the truncation error across the entire domain below a predefined threshold. This can be achieved by either increasing the expansion order $k$ or, for a fixed order, by reducing the size of the domain for a single expansion and patching several polynomials to cover the initial domain. Since the number of polynomial coefficients grows exponentially with $k$, increasing the expansion order becomes computationally intractable above a certain threshold. The second idea was formalized in~\cite{Wittig2015a} with the development of an algorithm for the automatic control of the truncation error of Taylor expansions. This technique, called \glsentryfull{ads}, monitors the accuracy of $\mathcal{T}_{\vb*{y}}(\delta\vb*{x})$ by estimating the magnitude of the coefficients of order $k+1$ and splits the initial domains into two smaller ones as soon as the estimated coefficients grow above a predefined threshold. This operation is performed recursively for each subdomain until each map is deemed sufficiently accurate in its domain of interest. Starting from a single polynomial $[\vb*{x}]$ and the function $\vb*{f}$ to be evaluated, the procedure generates two sets of polynomials, or manifolds, for both the domain and its image through $\vb*{f}$, defined as
\begin{subequations}
\begin{align}
    M_{\vb*{x}} &= \left\{[\vb*{x}^{(i)}]:\bigcup_{i=1}^N [\vb*{x}^{(i)}]=[\vb*{x}] \right\}\\
    M_{\vb*{y}} &= \left\{[\vb*{y}^{(i)}]:\bigcup_{i=1}^N [\vb*{y}^{(i)}]=[\vb*{y}] \right\}
\end{align}
\label{eq:ads_man_def}
\end{subequations}
where $[\vb*{y}]=\vb*{f}([\vb*{x}])$ and $N$ is the total number of subdomains generated by the algorithm.

\subsection{Measurements regression}\label{sec:regression}

Ground sensors usually provide multiple measurements of the target \glsentryfull{so} taken at different epochs within the same observation window. However, the developed \glsentryshort{iod} algorithms use at most three instants to estimate the object's state. To maximize the information conveyed from the raw measurements to the \glsentryshort{iod} solution, the sensor data can be preprocessed as follows. Each observable is treated as an independent Gaussian random variable and polynomial regression is employed to estimate the observed quantities and the corresponding \glsentryfullpl{ci} at the epochs required by the \glsentryshort{iod} algorithms~\cite{Pirovano2020e,Losacco2023}.

Consider a set of $N$ measurements for the generic observable $Y$, where each entry is normally distributed and characterized by its mean value $y_i$ and standard deviation $\sigma_i$. This set is denoted as
\begin{equation}
    \left\{t_i;(y_i,\sigma_i)\right\} \qquad i\in[1,N]
    \label{eq:obs_set}
\end{equation}
with $t_i$ the observation epoch and $Y_i\sim\mathcal{N}(y_i,\sigma_i)$ the independent random variable at $t_i$. A \glsentryfull{ls} problem is then set up to fit the observed data and obtain the $m+1$ coefficients that model a polynomial dependency between the observation epoch and the observed measurements. The design matrix is firstly built as
\begin{equation}
    A=\begin{bmatrix}
        1 & t_1-t_0 & \ldots & (t_1-t_0)^m\\
        \vdots & \vdots & \ddots & \vdots\\
        1 & t_N-t_0 & \ldots & (t_N-t_0)^m\\
    \end{bmatrix}
\end{equation}
where $t_0$ is the regression epoch, selected as the epoch $t_i$ closest to the middle of the observation window, and $m\in[1,N-2]$ is the regression order. The \glsentryshort{ls} problem is then given by
\begin{equation}
    \vb*{y}=A\vb*{z}
    \label{eq:regression_ls}
\end{equation}
with $\vb*{y}=\{y_1,\ldots,y_n\}^T$ the real observations and $\vb*{z}=\{z_0,\ldots,z_m\}^T$ the regression parameters. If available, prior information on the measurement uncertainty can be exploited in the solution of \cref{eq:regression_ls} by introducing a weight matrix $W=\text{diag}(1/\sigma_i^2)$ where the $\sigma_i$ are the measurements' standard deviations. The solution to the \glsentrylong{wls} problem is then obtained as
\begin{equation}
    \hat{\vb*{z}} = (A^TWA)^{-1}A^TW\vb*{y}
    \label{eq:regression_wls}
\end{equation}
whereas the estimated measurements $\hat{\vb*{y}}$ are given by
\begin{equation}
    \hat{\vb*{y}}=A\hat{\vb*{z}}
    \label{eq:regressed_y}
\end{equation}
Finally, the covariance matrices of the estimated parameters $\hat{\vb*{z}}$ and measurements $\hat{\vb*{y}}$ are computed as
\begin{subequations}
    \begin{align}
        P_{\hat{\vb*{z}}\hat{\vb*{z}}} &= (A^TWA)^{-1}\\
        P_{\hat{\vb*{y}}\hat{\vb*{y}}} &= AP_{\hat{\vb*{z}}\hat{\vb*{z}}}A^T
        \label{eq:cov_estim_obs}
    \end{align}
\end{subequations}
If no prior information on the measurement uncertainty is available, the solution is obtained as
\begin{equation}
    \hat{\vb*{z}} = (A^TA)^{-1}A^T\vb*{y}
\end{equation}
while the estimated measurements $\hat{\vb*{y}}$ are still given by \cref{eq:regressed_y}. An estimate of the measurement uncertainty is then computed as
\begin{equation}
    \hat{\sigma}^2 = \dfrac{\hat{\vb*{r}}^T\hat{\vb*{r}}}{N-p}
\end{equation}
with $\hat{\vb*{r}}=\vb*{y}-\hat{\vb*{y}}$ the measurement residuals and $p=m+1$. The covariance matrix of the fitted parameters is obtained as
\begin{equation}
    P_{\hat{\vb*{z}}\hat{\vb*{z}}} = \hat{\sigma}^2(A^TA)^{-1}
\end{equation}
whereas $P_{\hat{\vb*{y}}\hat{\vb*{y}}}$ is given by \cref{eq:cov_estim_obs}. If the hypothesis of independence between measurements holds, it can be shown that
\begin{equation}
    \dfrac{\hat{y}_i-y_i}{\sqrt{P_{\hat{\vb*{y}}\hat{\vb*{y}},ii}}} \sim t_{N-p}
\end{equation}
where $t_{N-p}$ denotes the Student's t-distribution with $N-p$ \glsentrylongpl{dof}. An estimate for the \glsentryshort{ci} of the $i^{th}$ observation is then obtained as
\begin{equation}
    \text{CI}_{\hat{y}_i} = \left[\hat{y}_i \pm q_{\frac{1+\alpha}{2},N-p}\cdot\sqrt{P_{\hat{\vb*{y}}\hat{\vb*{y}},ii}}\right] = \left[\hat{y}_i \pm \Delta\text{CI}_{\hat{y}_i}\right]
    \label{eq:ci_def}
\end{equation}
where $\alpha$ is the \glsentrylong{cl} and $q_{\frac{1+\alpha}{2},N-p}$ is the quantile function (or inverse \glsentrylong{cdf}) of $t_{N-p}$ evaluated at $\frac{1+\alpha}{2}$. The \glsentryshort{ci} is thus the interval within which the true value can be found with a \glsentrylong{cl} $\alpha$.

The developed \glsentryshort{iod} algorithm requires that the input measurements are initialized as \glsentryshort{da} variables to obtain a polynomial representation of their nominal values and associated uncertainties. In the \glsentryshort{da} framework these are denoted as
\begin{equation}
    [y_i] = \bar{y}_i + \beta_{y_i}\delta y_i
    \label{eq:da_init}
\end{equation}
where $\bar{y}_i$ is the nominal value, $\delta y_i$ the first-order deviation in $y_i$ and $\beta_{y_i}\in\mathbb{R}_{\geq 0}$ a scaling coefficient for the measurement uncertainty. If measurements regression is performed, \cref{eq:da_init} is rewritten as
\begin{equation}
    [y_i] = \hat{y}_i + \Delta\text{CI}_{\hat{y}_i}\cdot\delta y_i
    \label{eq:da_init_ci}
\end{equation}
where $\hat{y}_i$ and $\Delta\text{CI}_{\hat{y}_i}$ are the estimated measurements and the associated \glsentrylongpl{ci} given by \cref{eq:regressed_y,eq:ci_def}, respectively. These values are the result of polynomial regression and are thus dependent on the control parameters $m$ and $\alpha$. Lower regression orders $m$ and larger \glsentrylongpl{cl} $\alpha$ result in larger $\Delta\text{CI}_{\hat{y}_i}$ which are more likely to include the true, yet unknown, measurement. At the same time, when solving the \glsentryshort{iod} problem within the \glsentryshort{ads} framework, larger \glsentrylongpl{ci} may trigger more splits. The two parameters have thus to be chosen for the best tradeoff between accuracy, quantified by the likelihood of including the true measurements within the \glsentryshort{da} variables' bounds, and computational effort, quantified by the number of domains generated by the \glsentryshort{ads} algorithm.

If raw observations are used instead, $[y_i]$ is initialized as
\begin{equation}
    [y_i] = y_i + 3\sigma_i\cdot\delta y_i
    \label{eq:da_init_3s}
\end{equation}
where $y_i$ and $\sigma_i$ are the mean and standard deviation of the measurement as given by \cref{eq:obs_set}.

\subsection{Dynamical models}\label{sec:dynamics}

The \glsentryshort{iod} algorithms developed in \cref{sec:j2_pert_iod} start by computing a nominal solution to the \glsentryshort{iod} problem under the assumption of unperturbed Keplerian motion. The latter is then refined in a higher fidelity dynamical model to compensate for deviations from the nominal two-body trajectory, which may be non-negligible for long observation windows, and to compute a polynomial expansion of the solution with respect to the measurement uncertainty. The proposed algorithm is agnostic with respect to the dynamical model used for the refinement, provided that the solution remains sufficiently close to the initial guess. However, since a computationally efficient propagator is desirable at this stage, the analytical formulation of the $J_2$-perturbed dynamics proposed in~\cite{Armellin2018h} is used for all numerical simulations presented in \cref{sec:num_simu}.

\section{Perturbed \texorpdfstring{\glsentrylong{iod}}{Initial orbit determination}}\label{sec:j2_pert_iod}
This section describes the core of the \glsentryshort{iod} algorithms, namely the computation of a Taylor expansion of the \glsentryshort{so}'s state at the \glsentryshort{iod} epoch as a function of the uncertainty on the input measurements. These computations are wrapped within the \glsentryshort{ads} algorithm introduced in \cref{sec:ads} to control the accuracy of the final solution. If no split is triggered, the object state is represented as a single polynomial $[\vb*{x}]$. Instead, if one or more splits are required to satisfy the threshold imposed on the truncation error, the same state is described by a manifold of polynomials $M_{\vb*{x}}$ as defined in \cref{eq:ads_man_def}. A solution to the \glsentryshort{iod} problem is presented for the three types of ground-based sensors: range radars (\cref{sec:rr}), Doppler-only radars (\cref{sec:dr}), and optical telescopes (\cref{sec:o}).

\subsection{Range radars}
\label{sec:rr}
\begin{figure}[h!]
    \centering
    \includegraphics[trim={1.8cm 1.5cm 1.8cm 1.3cm},clip,width=0.5\textwidth]{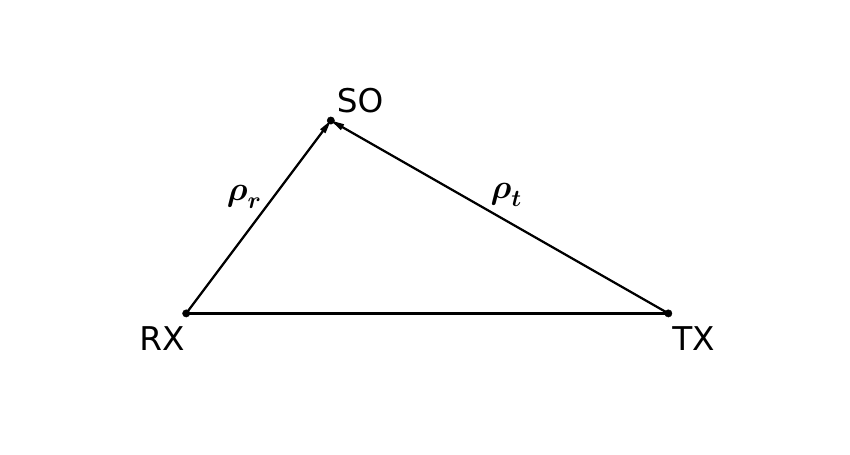}
    \caption{Geometry of bistatic radar sensor}
    \label{fig:bistatic_radar}
\end{figure}

Consider a range radar whose \glsentryfull{rx} and \glsentryfull{tx} are identified by the geodetic coordinates $(\phi_r, \lambda_r, h_r)$ and $(\phi_t,\lambda_t,h_t)$, with $\phi_{r,t}\in[-\pi/2,\pi/2]$ geodetic latitude, $\lambda_{r,t}\in[-\pi,\pi]$ geodetic longitude and $h_{r,t}\in\mathbb{R}_{\geq 0}$ geodetic height of receiver and transmitter, respectively. Three observables are provided for each detection instant. The first two measurements, azimuth $\vartheta$ and elevation $\varphi$, define the angular position of the \glsentryfull{so} in the receiver topocentric reference frame. The former provides the angular displacement of the object from the North, measured eastward on the local horizon, such that $\vartheta\in[0,2\pi]$. The latter is the angle between the line of sight and the horizon, $\varphi\in[-\pi/2,\pi/2]$. The third observable is the slant range $d=\rho_{r}+\rho_{t}$, that is, the sum of the distances of the object from the receiver and transmitter, as shown in \cref{fig:bistatic_radar}.

Denote with $\vb*{x}=[\vb*{r}^T\ \vb*{v}^T]^T$ the inertial state vector of the tracked \glsentryshort{so} and with $\vb*{r}_{r}$ the inertial position of the radar receiver at the same epoch. The relationship between the two position vectors $\vb*{r}$ and $\vb*{r}_{r}$ can be expressed as
\begin{equation}
\label{eq:rr_r}
    \vb*{r} = \vb*{r}_{r}+\rho\vb*{R}\vb*{\rho}_u
\end{equation}
The unit vector $\vb*{\rho}_u$ is the line of sight expressed in the receiver topocentric frame, i.e.
\begin{equation}
\label{eq:rr_los}
    \vb*{\rho}_u = \begin{bmatrix}
        \cos{\vartheta}\cos{\varphi}\\
        -\sin{\vartheta}\cos{\varphi}\\
        \sin{\varphi}
    \end{bmatrix}
\end{equation}
The matrix $\vb*{R}$ is a time-dependent rotational matrix that converts $\vb*{\rho}_u$ from the receiver topocentric to the inertial frame~\cite{Vallado2013a}, whereas $\rho=\rho_r$ is the object range as measured from the receiver. The latter is not directly available but can be retrieved by knowing the angular position of the object, the slant range, and the location of receiver and transmitter: $\rho = \rho(\vartheta,\varphi,d,\vb*{r}_r,\vb*{r}_t)$~\cite{Yanez2017}.

\subsubsection{\texorpdfstring{\Glsentrylong{iod}}{Initial orbit determination}}
\label{sec:rr_iod}

Consider a set of $N$ tuples of angular and slant range measurements provided by a ground-based range radar while observing an unknown object
\begin{equation}
    \left\{t_i;\left(\vartheta_i;\sigma_{\vartheta_i}\right),\left(\varphi_i;\sigma_{\varphi_i}\right),\left(d_i;\sigma_{d_i}\right)\right\}\qquad i\in[1,N]
    \label{eq:rr_meas_set}
\end{equation}
with $\vartheta_i,\varphi_i,d_i$ azimuth, elevation and slant range of the object at epoch $t_i$ and $\sigma_{\alpha_i},\sigma_{\delta_i},\sigma_{d_i}$ the associated standard deviations of the sensor noise, assumed as uncorrelated white noise. The approach presented here extends the method proposed by~\cite{Armellin2018} to perturbed orbital dynamics and provides a mathematical expression for the uncertainty on the state estimate as a function of the deviations in the nominal measurements. The first step involves a preliminary estimate of the full orbital state at epoch $t_1$. Knowing the angular position and the slant ranges at the first and final epochs, here defined as $t_1$ and $t_2$, the inertial positions of the object $\vb*{r}_1$ and $\vb*{r}_2$ are automatically known by applying geometrical considerations~\cite{Yanez2017} and \cref{eq:rr_r,eq:rr_los}. An estimate of the velocities at the same epochs is then obtained by solving the Lambert's problem~\cite{Izzo2015} in $[t_1,t_2]$ as
\begin{equation}
\label{eq:rr_lambert}
\mathcal{L}\left(t_1,\vb*{r}_1, t_2, \vb*{r}_2\right)\rightarrow \left\{\vb*{v}_{1,\mathcal{K}},\vb*{v}_{2,\mathcal{K}}\right\}
\end{equation}
where the subscript $\mathcal{K}$ indicates that the two velocities were obtained under the hypothesis of Keplerian dynamics. Once an estimate of the orbital state at $t_1$ is available, a correction is computed to account for the perturbed dynamics and the uncertainty on the measurements. The angular and range measurements at $t_1$ and $t_2$ are initialized as \glsentryshort{da} variables 
\begin{equation}
\label{eq:rr_da_meas}
    \begin{aligned}
    \left[\vartheta_j\right] &= \bar{\vartheta}_j+\beta_{\vartheta_j}\delta\vartheta_j\\
    \left[\varphi_j\right] &= \bar{\varphi}_j+\beta_{\varphi_j}\delta\varphi_j\\
    \left[d_j\right] &= \bar{d}_j+\beta_{d_j}\delta d_j\\
    \end{aligned}
\end{equation}
where $\bar{\vartheta}_j,\bar{\varphi}_j,\bar{d}_j$ are the nominal measurements and $\beta_{\vartheta_j},\beta_{\varphi_j},\beta_{d_j}$ are the associated uncertainty scaling factors given by \cref{eq:da_init_ci} or \cref{eq:da_init_3s} for $j=1,2$. The Taylor expansions of the inertial positions of the object at the two epochs can then be computed as
\begin{equation}
\label{eq:rr_da_pos}
\left[\vb*{r}_j\right]=\mathcal{T}_{\vb*{r}_j}\left(\delta\vartheta_j,\delta\varphi_j,\delta d_j\right)  
\end{equation}
By considering three additional \glsentryshort{da} variables, the expansion of the velocity at $t_1$ can be expressed as
\begin{equation}
    \left[\vb*{v}_1\right] = \vb*{v}_{1,\mathcal{K}}+\delta\vb*{v}_1 = \mathcal{T}_{\vb*{v}_1}\left(\delta\vb*{v}_1\right)
    \label{eq:v1_lamb_da}
\end{equation}
where $\vb*{v}_{1,\mathcal{K}}$ is given by \cref{eq:rr_lambert} and $\delta\vb*{v}_1=\left\{\delta v_{1,x},\delta v_{1,y},\delta v_{1,z}\right\}$. Consequently, the Taylor expansion of the orbital state at epoch $t_1$ can be expressed as
\begin{equation}
\label{eq:rr_DA_state_t1}
    \left[\vb*{x}_1\right]=
    \begin{Bmatrix}
        \left[\vb*{r}_1\right]\\
        \left[\vb*{v}_1\right]
    \end{Bmatrix}=
    \begin{Bmatrix}
        \mathcal{T}_{\vb*{r}_1}\left(\delta\vartheta_1,\delta\varphi_1,\delta d_1\right)\\
        \mathcal{T}_{\vb*{v}_1}\left(\delta\vb*{v}_1\right)
    \end{Bmatrix} = \mathcal{T}_{\vb*{x}_1}\left(\delta\vartheta_1,\delta\varphi_1,\delta d_1,\delta\vb*{v}_1\right)
\end{equation}
The orbital state $\left[\vb*{x}_1\right]$ is then propagated to $t_2$ under perturbed dynamics to obtain the estimated state at $t_2$
\begin{equation}
\label{eq:rr_DA_state_t2_prop}
    \left[\hat{\vb*{x}}_{2}\right]=
    \begin{Bmatrix}
        \left[\hat{\vb*{r}}_{2}\right]\\
        \left[\hat{\vb*{v}}_{2}\right]
    \end{Bmatrix}=
    \mathcal{T}_{\hat{\vb*{x}}_{2}}\left(\delta\vartheta_1,\delta\varphi_1,\delta d_1,\delta\vb*{v}_1\right)
\end{equation}
The hat symbol indicates that the quantity is an estimate resulting from the perturbed propagation process, and is used whenever the perturbed dynamics come into play. At this point, the difference between the inertial position vectors at $t_2$ as obtained from \cref{eq:rr_da_pos,eq:rr_DA_state_t2_prop} is expressed as
\begin{equation}
    \begin{aligned}
        \left[\Delta\hat{\vb*{r}}_2\right] = \left[\hat{\vb*{r}}_{2}\right]-\left[\vb*{r}_2\right] &= \mathcal{T}_{\hat{\vb*{r}}_{2}}\left(\delta\vartheta_1,\delta\varphi_1,\delta d_1,\delta\vb*{v}_1\right)-\mathcal{T}_{\vb*{r}_{2}}\left(\delta\vartheta_2,\delta\varphi_2,\delta d_2\right)\\
        &= \mathcal{T}_{\Delta\hat{\vb*{r}}_{2}}\left(\delta\vb*{\vartheta},\delta\vb*{\varphi},\delta\vb*{d},\delta\vb*{v}_1\right)
    \end{aligned}
    \label{eq:rr_delta_r_t2}
\end{equation}
where $\delta\vb*{\vartheta}=\left\{\delta\vartheta_1,\delta\vartheta_2\right\}$, $\delta\vb*{\varphi}=\left\{\delta\varphi_1,\delta\varphi_2\right\}$ and $\delta\vb*{d}=\left\{\delta d_1,\delta d_2\right\}$. \Cref{eq:rr_delta_r_t2} is then split into its constant part and polynomial expansion, namely
\begin{equation}
\label{eq:rr_delta_r_t2_2}
\left[\Delta\hat{\vb*{r}}_2\right]=\Delta\hat{\vb*{r}}_2+ \mathcal{T}_{\delta\hat{\vb*{r}}_{2}}\left(\delta\vb*{\vartheta},\delta\vb*{\varphi},\delta\vb*{d},\delta\vb*{v}_1\right)
\end{equation}
an augmented map is then built as
\begin{equation}
\label{eq:rr_map}
    \begin{Bmatrix}
        \delta\vb*{\vartheta}\\
        \delta\vb*{\varphi}\\
        \delta\vb*{d}\\
        \delta\hat{\vb*{r}}_2\\
    \end{Bmatrix}=
    \begin{Bmatrix}
        \delta\vb*{\vartheta}\\
        \delta\vb*{\varphi}\\
        \delta\vb*{d}\\
        \mathcal{T}_{\delta\hat{\vb*{r}}_{2}}\left(\delta\vb*{\vartheta},\delta\vb*{\varphi},\delta\vb*{d},\delta\vb*{v}_1\right)\\
    \end{Bmatrix}
\end{equation}
and map inversion is exploited to obtain
\begin{equation}
\label{eq:rr_da_dv1}
    \delta\vb*{v}_1 = \mathcal{T}_{\delta\vb*{v}_1}\left(\delta\vb*{\vartheta},\delta\vb*{\varphi},\delta\vb*{d},\delta\hat{\vb*{r}}_2\right)
\end{equation}
The dependency of $\left[\vb*{x}_1\right]$ on $\delta\vb*{v}_1$ is then removed by composing \cref{eq:rr_DA_state_t1} with \cref{eq:rr_da_dv1}
\begin{equation}
\begin{aligned}
\label{eq:rr_da_state_t1_comp}
    \left[\hat{\vb*{x}}_1\right]&=
    \mathcal{T}_{\vb*{x}_1}\left(\delta\vartheta_1,\delta\varphi_1,\delta d_1,\delta\vb*{v}_1\right)\circ \mathcal{T}_{\delta\vb*{v}_1}\left(\delta\vb*{\vartheta},\delta\vb*{\varphi},\delta\vb*{d},\delta\hat{\vb*{r}}_2\right) \\
    &=\mathcal{T}_{\hat{\vb*{x}}_1}\left(\delta\vb*{\vartheta},\delta\vb*{\varphi},\delta\vb*{d},\delta\hat{\vb*{r}}_2\right)
\end{aligned} 
\end{equation}
The continuity of the inertial position of the object at $t_2$ can finally be enforced by evaluating \cref{eq:rr_da_state_t1_comp} into $\left(\delta\vb*{\vartheta},\delta\vb*{\varphi},\delta\vb*{d},-\Delta\hat{\vb*{r}}_2\right)$ to obtain
\begin{equation}
\begin{aligned}
    \left[\hat{\vb*{x}}_1\right]&=
    \mathcal{T}_{\hat{\vb*{x}}_1}\left(\delta\vb*{\vartheta},\delta\vb*{\varphi},\delta\vb*{d},\delta\hat{\vb*{r}}_2\right)\circ\left(\delta\vb*{\vartheta},\delta\vb*{\varphi},\delta\vb*{d},-\Delta\hat{\vb*{r}}_2\right)\\
    &=\mathcal{T}_{\hat{\vb*{x}}_1}\left(\delta\vb*{\vartheta},\delta\vb*{\varphi},\delta\vb*{d}\right)
\end{aligned}
\label{eq:rr_da_state_t1_comp_eval}
\end{equation}
As a result, the Taylor expansion of the orbital state at epoch $t_1$ expressed in terms of the deviations from the nominal measurements is obtained.

\subsection{Doppler-only radars}
\label{sec:dr}
Consider a Doppler-only radar  whose receiver and transmitter are identified by the geodetic coordinates $(\phi_r,\lambda_r,h_r)$ and $(\phi_t,\lambda_t,h_t)$, with $\phi_{r,t}\in[-\pi/2,\pi/2]$ geodetic latitude, $\lambda_{r,t}\in[-\pi,\pi]$ geodetic longitude and $h_{r,t}\in\mathbb{R}_{\geq 0}$ geodetic height of receiver and transmitter, respectively. Three observables are provided at each detection epoch. The first two are the object azimuth and elevation as measured by the receiver (see \cref{sec:rr}). The third quantity is the range rate, defined as the time derivative $\dot{d}$ of the slant range $d$ introduced in \cref{sec:rr}. Now, let $\vb*{x}=[\vb*{r}^T\ \vb*{v}^T]^T$ be the inertial state vector of the tracked \glsentryshort{so}, and let $\vb*{r}_{r}$ be the inertial position of the radar receiver at the same epoch. The relationship between the two position vectors $\vb*{r}$ and $\vb*{r}_{r}$ is given by \cref{eq:rr_r}. However, the range $\rho$ is not available and has to be estimated as explained below.

\subsubsection{\texorpdfstring{\Glsentrylong{iod}}{Initial orbit determination}}
\label{sec:dr_iod}

Consider a set of $N$ tuples of angular and range rate measurements provided by a ground-based Doppler radar while observing an unknown object
\begin{equation}
\label{eq:dr_meas_set}
    \left\{t_i;\left(\vartheta_i;\sigma_{\vartheta_i}\right),\left(\varphi_i;\sigma_{\varphi_i}\right),\left(\dot{d}_i;\sigma_{\dot{d}_i}\right)\right\}\qquad i\in[1,N]
\end{equation}
with $\vartheta_i,\varphi_i,\dot{d}_i$ azimuth, elevation and range rate of the object at epoch $t_i$ and $\sigma_{\alpha_i},\sigma_{\delta_i},\sigma_{\dot{d}_i}$ the associated standard deviation of the sensor noise, assumed as uncorrelated white noise. Like the range radar case, \glsentryshort{da} techniques can be exploited to estimate both the nominal state and the uncertainty of the transiting object. However, unlike the previous case, the lack of range information introduces additional challenges to be faced while estimating the orbital state. The approach presented here is an extension of the algorithm proposed by~\cite{Losacco2023} and runs in two phases: range estimation and state estimate expansion.

\paragraph{Range estimation}
\label{sec:dr_iod_range_est}
Consider the sets of angular measurements at the first ($t_1$), middle ($t_2$), and last ($t_3$) observation epochs, i.e.
$\left(\vartheta_1;\varphi_1\right),\left(\vartheta_2;\varphi_2\right),\left(\vartheta_3;\varphi_3\right)$. A Gauss' problem can be set up, thus obtaining an estimate of the object ranges at these epochs
\begin{equation}
\mathcal{G}\left(t_1,\vartheta_1,\varphi_1,t_2,\vartheta_2,\varphi_2,t_3,\vartheta_3,\varphi_3\right)\rightarrow\left\{\rho_{1,\mathcal{K}},\rho_{2,\mathcal{K}},\rho_{3,\mathcal{K}}\right\}
\end{equation}
where the subscript $\mathcal{K}$ indicates that Keplerian dynamics is assumed. The ranges at the first and last epochs are then initialized as \glsentryshort{da} variables, namely
\begin{equation}
\label{eq:dr1_da_rho}
\left[\rho_j\right] = \rho_{j,\mathcal{K}}+\delta\rho_j
\end{equation}
with $j=1,3$. As a result, the inertial positions at $t_1$ and $t_3$ can be built using \cref{eq:rr_r,eq:rr_los}, thus obtaining
\begin{equation}
\label{eq:dr1_da_rr}
\left[\vb*{r}_j\right] = \mathcal{T}_{\vb*{r}_j}\left(\delta\rho_j\right)
\end{equation}
An estimate for the velocity vector at $t_1$ is then obtained by solving the Lambert's problem in $[t_1,t_3]$ and initializing the output as a \glsentryshort{da} variable as in \cref{eq:v1_lamb_da}. The six-dimensional orbital state is then built as
\begin{equation}
\label{eq:dr1_da_x1}
    \left[\vb*{x}_1\right]=
    \begin{Bmatrix}
    \left[\vb*{r}_1\right]\\
    \left[\vb*{v}_1\right]
    \end{Bmatrix}=
    \begin{Bmatrix}
    \mathcal{T}_{\vb*{r}_1}\left(\delta\rho_1\right)\\
    \vb*{v}_{1,\mathcal{K}}+\delta\vb*{v}_1
    \end{Bmatrix}=
    \mathcal{T}_{\vb*{x}_1}\left(\delta\rho_1,\delta\vb*{v}_1\right)
\end{equation}
Propagating $\left[\vb*{x}_1\right]$ to $t_3$ under perturbed dynamics then leads to
\begin{equation}
\label{eq:dr1_da_x3}
    \left[\hat{\vb*{x}}_{3}\right] = \mathcal{T}_{\hat{\vb*{x}}_{3}}\left(\delta\rho_1,\delta\vb*{v}_1\right)
\end{equation}
The position residuals at $t_3$ are then obtained from \cref{eq:dr1_da_x3,eq:dr1_da_rr}
\begin{equation}
\label{eq:dr1_da_delta_rr3}
\begin{aligned}
    \left[\Delta\hat{\vb*{r}}_3\right] = \left[\hat{\vb*{r}}_{3}\right]-\left[\vb*{r}_{3}\right] &= \mathcal{T}_{\hat{\vb*{r}}_{3}}\left(\delta\rho_1,\delta\vb*{v}_1\right)-\mathcal{T}_{\vb*{r}_{3}}\left(\delta\rho_3\right)\\
    &= \Delta\hat{\vb*{r}}_3+\mathcal{T}_{\delta\hat{\vb*{r}}_{3}}\left(\delta\vb*{\rho},\delta\vb*{v}_1\right)    
\end{aligned}
\end{equation}
where $\delta\vb*{\rho}=\left\{\delta\rho_1,\delta\rho_3\right\}$. An augmented map is then built as
\begin{equation}
\label{eq:dr1_map_rho}
\begin{Bmatrix}
    \delta\vb*{\rho}\\
    \delta\hat{\vb*{r}}_3
\end{Bmatrix}=
\begin{Bmatrix}
    \delta\vb*{\rho}\\
    \mathcal{T}_{\delta\hat{\vb*{r}}_{3}}\left(\delta\vb*{\rho},\delta\vb*{v}_1\right)
\end{Bmatrix}
\end{equation}
and map inversion can be exploited to obtain the Taylor expansion of $\delta\vb*{v}_1$ as a function of $\delta\vb*{\rho}$ and $\delta\hat{\vb*{r}}_3$, i.e.
\begin{equation}
\label{eq:dr1_dv1}
    \delta\vb*{v}_1 = \mathcal{T}_{\delta\vb*{v}_{1}}\left(\delta\vb*{\rho},\delta\hat{\vb*{r}}_3\right)
\end{equation}
The dependency of $\left[\vb*{x}_{1}\right]$ and $\left[\hat{\vb*{x}}_{3}\right]$ on $\delta\vb*{v}_{1}$ is eliminated by composing \cref{eq:dr1_da_x1,eq:dr1_da_x3} with \cref{eq:dr1_dv1}
\begin{equation}
\label{eq:dr1_da_x13_comp}
\begin{aligned}
    \left[\hat{\vb*{x}}_{1}\right] &= \mathcal{T}_{\vb*{x}_1}\left(\delta\rho_1,\delta\vb*{v}_1\right)\circ\mathcal{T}_{\delta\vb*{v}_{1}}\left(\delta\vb*{\rho},\delta\hat{\vb*{r}}_3\right) = \mathcal{T}_{\hat{\vb*{x}}_{1}}\left(\delta\vb*{\rho},\delta\hat{\vb*{r}}_3\right)\\
    \left[\hat{\vb*{x}}_{3}\right] &= \mathcal{T}_{\hat{\vb*{x}}_{3}}\left(\delta\rho_1,\delta\vb*{v}_1\right)\circ\mathcal{T}_{\delta\vb*{v}_{1}}\left(\delta\vb*{\rho},\delta\hat{\vb*{r}}_3\right) = \mathcal{T}_{\hat{\vb*{x}}_{3}}\left(\delta\vb*{\rho},\delta\hat{\vb*{r}}_3\right)
\end{aligned}    
\end{equation}
Finally, the continuity of the position vectors at $t_3$ is enforced by evaluating \cref{eq:dr1_da_x13_comp} in $\left(\delta\vb*{\rho},-\Delta\hat{\vb*{r}}_3\right)$ to obtain
\begin{equation}
\label{eq:dr1_da_x13_comp_eval}
\begin{aligned}
    \left[\hat{\vb*{x}}_{1}\right] &= \mathcal{T}_{\hat{\vb*{x}}_1}\left(\delta\vb*{\rho},\delta\hat{\vb*{r}}_3\right)\circ\left(\delta\vb*{\rho},-\Delta\hat{\vb*{r}}_3\right) = \mathcal{T}_{\hat{\vb*{x}}_{1}}\left(\delta\vb*{\rho}\right)\\
    \left[\hat{\vb*{x}}_{3}\right] &= \mathcal{T}_{\hat{\vb*{x}}_{3}}\left(\delta\vb*{\rho},\delta\hat{\vb*{r}}_3\right)\circ\left(\delta\vb*{\rho},-\Delta\hat{\vb*{r}}_3\right) = \mathcal{T}_{\hat{\vb*{x}}_{3}}\left(\delta\vb*{\rho}\right)\\
\end{aligned}    
\end{equation}
The described process removes the dependency on $\delta\vb*{v}_1$ but does not resolve the ambiguity on the range variables, which must be solved for. This is solved by imposing a match between the estimated and measured range rates at $t_1$ and $t_3$. Specifically, starting from \cref{eq:dr1_da_x13_comp_eval}, the estimate of the Taylor expansion of the range rates at the two epochs can be written as
\begin{equation}
\label{eq:dr1_doppler}
    \left[\hat{\dot{d}}_j\right] = \mathcal{T}_{\hat{\dot{d}}_j}(\delta\vb*{\rho})
\end{equation}
for $j=1,3$. The difference between the estimated and measured rates is then expressed as
\begin{equation}
    \begin{aligned}
        \left[\Delta\hat{\dot{\vb*{d}}}\right]&=\left[\hat{\dot{\vb*{d}}}\right]-\dot{\vb*{d}}\\
        &=\Delta\hat{\dot{\vb*{d}}}+\mathcal{T}_{\delta\hat{\dot{\vb*{d}}}}(\delta\vb*{\rho})
    \end{aligned}
\end{equation}
with $\left[\hat{\dot{\vb*{d}}}\right]=\left\{\left[\hat{\dot{d}}_1\right],\left[\hat{\dot{d}}_3\right]\right\}$ built from \cref{eq:dr1_doppler} and $\dot{\vb*{d}}=\{\dot{d}_1,\dot{d}_3\}$ measured range rates.
As a result, the map $\mathcal{T}_{\delta\hat{\dot{\vb*{d}}}}(\delta\vb*{\rho})$ can be inverted to obtain the dependency of $\delta\vb*{\rho}$ on $\delta\hat{\dot{\vb*{d}}}$, that is
\begin{equation}
\label{eq:dr1_da_drho}
    \delta\vb*{\rho} = \mathcal{T}_{\delta\vb*{\rho}}\left(\delta\hat{\dot{\vb*{d}}}\right)
\end{equation}
The matching between range rates is finally enforced by evaluating \cref{eq:dr1_da_drho} in $-\Delta\hat{\dot{\vb*{d}}}$, that is
\begin{equation}
\label{eq:dr1_delta_rho}
    \Delta\vb*{\rho} = \mathcal{T}_{\delta\vb*{\rho}}\left(-\Delta\hat{\dot{\vb*{d}}}\right)
\end{equation}
The quantity $\Delta\vb*{\rho}$ represents the correction to the constant part of \cref{eq:dr1_da_rho}. Once computed, the process from \cref{eq:dr1_da_rho} to \cref{eq:dr1_delta_rho} is repeated until the correction on $\vb*{\rho}$ is below a predefined threshold, i.e. $\norm{\Delta\vb*{\rho}}<\varepsilon$.

\paragraph{State estimate expansion}
\label{sec:dr_iod_state_exp}

The process described in \cref{sec:dr_iod_range_est} provides an estimate of the object range at epochs $t_1$ and $t_3$, which can be used to estimate the orbital state by using \cref{eq:dr1_da_x13_comp_eval}. However, no information regarding the uncertainty of the estimate is available at this stage. As in \cref{sec:rr_iod}, a process that exploits the measurement accuracy is thus set up to estimate the uncertainty on the computed solution.

Consider the available measurements at epochs $t_1$ and $t_3$ and initialize them as \glsentryshort{da} variables
\begin{equation}
\label{eq:dr2_da_meas}
\begin{aligned}
    \left[\vartheta_j\right] &= \bar{\vartheta}_j+\beta_{\vartheta_j}\delta\vartheta_j\\
    \left[\varphi_j\right] &= \bar{\varphi}_j+\beta_{\varphi_j}\delta\varphi_j\\
    \left[\dot{d}_j\right] &= \bar{\dot{d}}_j+\beta_{\dot{d}_j}\delta\dot{d}_j
\end{aligned}    
\end{equation}
with $\bar{\vartheta}_j,\bar{\varphi}_j,\bar{\dot{d}}_j$ the nominal measurements and $\beta_{\vartheta_j},\beta_{\varphi_j},\beta_{\dot{d}_j}$ the uncertainty scaling factors given by \cref{eq:da_init_ci} or \cref{eq:da_init_3s} for $j=1,3$. Similarly, the estimated ranges can be expressed as
\begin{equation}
\label{eq:dr2_da_rho}
\left[\rho_j\right] = \rho_j+\delta\rho_j
\end{equation}
where the $\rho_j$ are the result of the estimation process described in \cref{sec:dr_iod_range_est}. Given the lines of sight and the ranges, the Taylor expansion of the inertial position vectors is expressed as
\begin{equation}
\label{eq:dr2_da_rr}
\left[\vb*{r}_j\right]=\mathcal{T}_{\vb*{r}_j}\left(\delta\vartheta_j,\delta\varphi_j,\delta \rho_j\right) 
\end{equation}
The Taylor expansion of the orbital state at epoch $t_1$ is then given by
\begin{equation}
\label{eq:dr2_x1}
    \left[\vb*{x}_1\right] = 
    \begin{Bmatrix}
    \left[\vb*{r}_1\right]\\
    \left[\vb*{v}_1\right]
    \end{Bmatrix}=\\
    \begin{Bmatrix}
    \mathcal{T}_{\vb*{r}_1}\left(\delta\vartheta_1,\delta\varphi_1,\delta\rho_1\right)\\
    \vb*{v}_{1}+\delta\vb*{v}_1
    \end{Bmatrix}=\\
    \mathcal{T}_{\vb*{x}_1}\left(\delta\vartheta_1,\delta\varphi_1,\delta\rho_1,\delta\vb*{v}_1\right)
\end{equation}
where $\vb*{v}_{1}$ becomes available after the range estimation phase. The state is then propagated to $t_3$ under perturbed dynamics, thereby obtaining
\begin{equation}
\label{eq:dr2_da_x3}
    \left[\hat{\vb*{x}}_{3}\right]=\mathcal{T}_{\hat{\vb*{x}}_{3}}\left(\delta\vartheta_1,\delta\varphi_1,\delta\rho_1,\delta\vb*{v}_1\right)
\end{equation}
The position residuals at $t_3$ are then obtained from \cref{eq:dr2_da_x3,eq:dr2_da_rr} as
\begin{equation}
\label{eq:dr2_da_delta_rr3}
    \left[\Delta\hat{\vb*{r}}_3\right] = \left[\hat{\vb*{r}}_{3}\right]-\left[\vb*{r}_{3}\right] = \Delta\hat{\vb*{r}}_3 +\mathcal{T}_{\delta\hat{\vb*{r}}_3}\left(\delta\vb*{\vartheta},\delta\vb*{\varphi},\delta\vb*{\rho},\delta\vb*{v}_1\right)
\end{equation}
An augmented map is then built as
\begin{equation}
\label{eq:dr2_map_dv}
    \begin{Bmatrix}
        \delta\vb*{\vartheta}\\
        \delta\vb*{\varphi}\\
        \delta\vb*{\rho}\\
        \delta\hat{\vb*{r}}_3\\
    \end{Bmatrix}=
    \begin{Bmatrix}
        \delta\vb*{\vartheta}\\
        \delta\vb*{\varphi}\\
        \delta\vb*{\rho}\\
        \mathcal{T}_{\delta\hat{\vb*{r}}_{3}}\left(\delta\vb*{\vartheta},\delta\vb*{\varphi},\delta\vb*{\rho},\delta\vb*{v}_1\right)\\
    \end{Bmatrix}
\end{equation}
and map inversion is exploited to obtain
\begin{equation}
\label{eq:dr2_da_dv1}
    \delta\vb*{v}_1 = \mathcal{T}_{\delta\vb*{v}_1}\left(\delta\vb*{\vartheta},\delta\vb*{\varphi},\delta\vb*{\rho},\delta\hat{\vb*{r}}_3\right)
\end{equation}
By composing \cref{eq:dr2_x1} and~\eqref{eq:dr2_da_x3} with \cref{eq:dr2_da_dv1}, and then evaluating the resulting polynomials in $\left(\delta\vb*{\vartheta},\delta\vb*{\varphi},\delta\vb*{\rho},-\Delta\hat{\vb*{r}}_3\right)$, the following expansions are obtained
\begin{equation}
\begin{aligned}
\label{eq:dr2_da_x13_comp_eval}
    \left[\hat{\vb*{x}}_{1}\right]&=
    \mathcal{T}_{\hat{\vb*{x}}_{1}}\left(\delta\vb*{\vartheta},\delta\vb*{\varphi},\delta\vb*{\rho}\right)\\
    \left[\hat{\vb*{x}}_{3}\right]&=
    \mathcal{T}_{\hat{\vb*{x}}_{3}}\left(\delta\vb*{\vartheta},\delta\vb*{\varphi},\delta\vb*{\rho}\right)
\end{aligned}   
\end{equation}
These expansions guarantee the matching of the position vectors at $t_3$ but do not consider the range rate measurements at the two epochs. A second map inversion is then used to enforce this matching and obtain a Taylor expansion function of the observables only. Specifically, starting from \cref{eq:dr2_da_x13_comp_eval}, the estimated range rates at epochs $t_1$ and $t_3$ can be computed as
\begin{equation}
\label{eq:dr2_doppler_j}
    \left[\hat{\dot{d}}_j\right] = \mathcal{T}_{\hat{\dot{d}}_j}(\delta\vb*{\vartheta},\delta\vb*{\varphi},\delta\vb*{\rho})
\end{equation}
The difference between the estimated and measured range rates at the two epochs is then expressed as
\begin{equation}
\begin{aligned}
\label{eq:dr2_delta_doppler_j}  
    \left[\Delta\hat{\dot{d}}_j\right] = \left[\hat{\dot{d}}_j\right]-\left[\dot{d}_j\right] &= \mathcal{T}_{\hat{\dot{d}}_j}(\delta\vb*{\vartheta},\delta\vb*{\varphi},\delta\vb*{\rho})-\mathcal{T}_{\dot{d}_j}(\delta\dot{d}_j)\\
    &=\mathcal{T}_{\Delta\hat{\dot{d}}_j}(\delta\vb*{\vartheta},\delta\vb*{\varphi},\delta\dot{d}_j,\delta\vb*{\rho})\\
    &=\Delta\hat{\dot{d}}_j+\mathcal{T}_{\delta\hat{\dot{d}}_j}(\delta\vb*{\vartheta},\delta\vb*{\varphi},\delta\dot{d}_j,\delta\vb*{\rho})
\end{aligned}
\end{equation}
By considering the residuals at both epochs, \cref{eq:dr2_delta_doppler_j} can be reformulated as
\begin{equation}
\begin{aligned}
    \label{eq:dr2_delta_doppler}  
    \left[\Delta\hat{\dot{\vb*{d}}}\right]&= 
    \begin{Bmatrix}
    \left[\Delta\hat{\dot{d}}_1\right]\\
    \left[\Delta\hat{\dot{d}}_3\right]
    \end{Bmatrix}= 
    \begin{Bmatrix}
    \Delta\hat{\dot{d}}_1+\mathcal{T}_{\delta\hat{\dot{d}}_1}(\delta\vb*{\vartheta},\delta\vb*{\varphi},\delta\dot{d}_1,\delta\vb*{\rho})\\
    \Delta\hat{\dot{d}}_3+\mathcal{T}_{\delta\hat{\dot{d}}_3}(\delta\vb*{\vartheta},\delta\vb*{\varphi},\delta\dot{d}_3,\delta\vb*{\rho})\\
    \end{Bmatrix}\\
    &=\Delta\hat{\dot{\vb*{d}}}+\mathcal{T}_{\delta\hat{\dot{\vb*{d}}}}(\delta\vb*{\vartheta},\delta\vb*{\varphi},\delta\dot{\vb*{d}},\delta\vb*{\rho})\\
\end{aligned}
\end{equation}
Subsequently, an augmented map is built
\begin{equation}
\label{eq:dr2_map}
    \begin{Bmatrix}
        \delta\vb*{\vartheta}\\
        \delta\vb*{\varphi}\\
        \delta\dot{\vb*{d}}\\
        \delta\hat{\dot{\vb*{d}}}\\
    \end{Bmatrix}=
    \begin{Bmatrix}
        \delta\vb*{\vartheta}\\
        \delta\vb*{\varphi}\\
        \delta\dot{\vb*{d}}\\
        \mathcal{T}_{\delta\hat{\dot{\vb*{d}}}}(\delta\vb*{\vartheta},\delta\vb*{\varphi},\delta\dot{\vb*{d}},\delta\vb*{\rho})\\
    \end{Bmatrix}
\end{equation}
from which map inversion yields
\begin{equation}
\label{eq:dr2_da_drho}
    \delta\vb*{\rho} = \mathcal{T}_{\delta\vb*{\rho}}\left(\delta\vb*{\vartheta},\delta\vb*{\varphi},\delta\dot{\vb*{d}},\delta\hat{\dot{\vb*{d}}}\right)
\end{equation}
The dependency of $\left[\hat{\vb*{x}}_j\right]$ on $\delta\vb*{\rho}$ is then eliminated by composing \cref{eq:dr2_da_x13_comp_eval} with \cref{eq:dr2_da_drho}
\begin{equation}
\label{eq:dr2_da_state_comp}
\begin{aligned}
    \left[\hat{\vb*{x}}_j\right]&=
    \mathcal{T}_{\hat{\vb*{x}}_j}\left(\delta\vb*{\vartheta},\delta\vb*{\varphi},\delta\vb*{\rho}\right)\circ \mathcal{T}_{\delta\vb*{\rho}}\left(\delta\vb*{\vartheta},\delta\vb*{\varphi},\delta\dot{\vb*{d}},\delta\hat{\dot{\vb*{d}}}\right)\\ 
    &=\mathcal{T}_{\hat{\vb*{x}}_j}\left(\delta\vb*{\vartheta},\delta\vb*{\varphi},\delta\dot{\vb*{d}},\delta\hat{\dot{\vb*{d}}}\right)
\end{aligned}    
\end{equation}
Finally, the Doppler shift measurements at $t_1$ and $t_3$ can be matched by evaluating \cref{eq:dr2_da_state_comp} in $\left(\delta\vb*{\vartheta},\delta\vb*{\varphi},\delta\dot{\vb*{d}},-\Delta\hat{\dot{\vb*{d}}}\right)$ to obtain
\begin{equation}
\begin{aligned}
\label{eq:dr2_da_state_comp_eval}
    \left[\hat{\vb*{x}}_j\right]&=
    \mathcal{T}_{\hat{\vb*{x}}_j}\left(\delta\vb*{\vartheta},\delta\vb*{\varphi},\delta\dot{\vb*{d}},\delta\hat{\dot{\vb*{d}}}\right)\circ\left(\delta\vb*{\vartheta},\delta\vb*{\varphi},\delta\dot{\vb*{d}},-\Delta\hat{\dot{\vb*{d}}}\right)\\
    &=\mathcal{T}_{\hat{\vb*{x}}_j}\left(\delta\vb*{\vartheta},\delta\vb*{\varphi},\delta\dot{\vb*{d}}\right)
\end{aligned}    
\end{equation}

\subsection{Optical telescopes}
\label{sec:o}
Consider an optical telescope identified by its geodetic coordinates $(\phi,\lambda,h)$ with $\phi\in[-\pi/2,\pi/2]$ geodetic latitude, $\lambda\in[-\pi,\pi]$ geodetic longitude and $h\in\mathbb{R}_{\geq 0}$ geodetic height. Denote with $\vb*{x}=[\vb*{r}^T\ \vb*{v}^T]^T$ the inertial state vector of the tracked object and with $\vb*{r}_{obs}$ the inertial position of the telescope at the same epoch. The two positions $\vb*{r}$ and $\vb*{r}_{obs}$ are related as follows
\begin{equation}
    \vb*{r} = \vb*{r}_{obs}+\rho\vb*{\rho}_u
\end{equation}
with $\rho$ topocentric range and $\vb*{\rho}_u$ \glsentryfull{los} unit vector. The latter is computed from the topocentric right ascension and declination $(\alpha,\delta)$ as
\begin{equation}
    \vb*{\rho}_u = \begin{bmatrix}
        \cos{\alpha}\cos{\delta}\\
        \sin{\alpha}\cos{\delta}\\
        \sin{\delta}
    \end{bmatrix}
    \label{eq:sph2los}
\end{equation}

\subsubsection{\texorpdfstring{\Glsentrylong{iod}}{Initial orbit determination}}
\label{sec:o_iod}

Consider a set of $N$ tuples of angular measurements provided by a ground-based optical sensor while observing an unknown \glsentryshort{so}
\begin{equation}
\label{eq:o_meas_set}
    \left\{t_i;\left(\alpha_i;\sigma_{\alpha_i}\right),\left(\delta_i;\sigma_{\delta_i}\right)\right\}\qquad i\in[1,N]
\end{equation}
with $\alpha_i,\delta_i$ the topocentric right ascension and declination of the \glsentryshort{so} at epochs $t_i$ and $\sigma_{\alpha_i},\sigma_{\delta_i}$ the associated standard deviations of the sensor noise, assumed as uncorrelated white noise. The proposed algorithm builds on~\cite{Pirovano2020e} to provide an estimate of the orbital solution and the associated uncertainty when considering perturbed dynamics. Similar to the Doppler radar case, this method consists of two phases as described hereafter.

\paragraph{Range estimation}
\label{sec:o_iod_range_est}
Consider the sets of angular measurements at the first ($t_1$), middle ($t_2$), and last ($t_3$) observation epochs, i.e. $\left(\alpha_1;\delta_1\right),\left(\alpha_2;\delta_2\right),\left(\alpha_3;\delta_3\right)$. A Gauss' problem is firstly solved to obtain an estimate of the object's ranges
\begin{equation}
\mathcal{G}\left(t_1,\alpha_1,\delta_1,t_2,\alpha_2,\delta_2,t_3,\alpha_3,\delta_3\right)\rightarrow\left\{\rho_{1,\mathcal{K}},\rho_{2,\mathcal{K}},\rho_{3,\mathcal{K}}\right\}
\end{equation}
The three ranges $\left[\rho_j\right]$ are then initialized as \glsentryshort{da} variables as in \cref{eq:dr1_da_rho} and an expression for the positions of the \glsentryshort{so} $\left[\vb*{r}_j\right]$ is computed similarly to \cref{eq:dr1_da_rr}. Following the procedure described in \cref{sec:dr_iod_range_est}, a Taylor expansion of the orbital state at $t_1$, $t_2$ and $t_3$ is then retrieved as in \cref{eq:dr1_da_x1}
\begin{equation}
    \left[\vb*{x}_j\right]=\mathcal{T}_{\vb*{x}_j}(\delta\rho_j,\delta\vb*{v}_j)
\end{equation}
where $j=1,2,3$. The state vector $\left[\vb*{x}_1\right]$ is then propagated to $t_2$ to obtain
\begin{equation}
\label{eq:o1_da_x2m}
    \left[\hat{\vb*{x}}^{-}_{2}\right] = \mathcal{T}_{\hat{\vb*{x}}^{-}_{2}}\left(\delta\rho_1,\delta\vb*{v}_1\right)
\end{equation}
where the negative sign indicates that the estimate is obtained starting from the state expansion at the earlier epoch $t_1$. Then, the Taylor expansion of the position residuals at $t_2$ is written as
\begin{equation}
\label{eq:o1_da_delta_rr2}
\begin{aligned}
    \left[\Delta\hat{\vb*{r}}_2^{-}\right] = \left[\hat{\vb*{r}}^{-}_{2}\right]-\left[\vb*{r}_{2}\right] &= \mathcal{T}_{\hat{\vb*{r}}^{-}_2}\left(\delta\rho_1,\delta\vb*{v}_1\right)-\mathcal{T}_{\vb*{r}_{2}}\left(\delta\rho_2\right)\\
    &=\Delta\hat{\vb*{r}}^{-}_2+\mathcal{T}_{\delta\hat{\vb*{r}}^{-}_{2}}\left(\delta\vb*{\rho}_{12},\delta\vb*{v}_1\right)    
\end{aligned}
\end{equation}
where $\delta\vb*{\rho}_{12} = \left\{\delta\rho_1,\delta\rho_2\right\}$. Subsequently, an augmented map is built
\begin{equation}
\label{eq:o1_map_rho1}
\begin{Bmatrix}
    \delta\vb*{\rho}_{12}\\
    \delta\hat{\vb*{r}}^{-}_2
\end{Bmatrix}=
\begin{Bmatrix}
    \delta\vb*{\rho}_{12}\\
    \mathcal{T}_{\delta\hat{\vb*{r}}^{-}_{2}}\left(\delta\vb*{\rho}_{12},\delta\vb*{v}_1\right)
\end{Bmatrix}
\end{equation}
and map inversion is used to obtain the Taylor expansion of $\delta\vb*{v}_1$, i.e.
\begin{equation}
\label{eq:o1_dv1}
    \delta\vb*{v}_1 = \mathcal{T}_{\delta\vb*{v}_{1}}\left(\delta\vb*{\rho}_{12},\delta\hat{\vb*{r}}^{-}_2\right)
\end{equation}
Composing this expression with the Taylor expansion of the orbital states at $t_1$ and $t_2$ yields
\begin{equation}
\label{eq:o1_da_x12m_comp}
\begin{aligned}
    \left[\hat{\vb*{x}}_{1}\right] &= \mathcal{T}_{\vb*{x}_1}\left(\delta\rho_1,\delta\vb*{v}_1\right)\circ\mathcal{T}_{\delta\vb*{v}_{1}}\left(\delta\vb*{\rho}_{12},\delta\hat{\vb*{r}}^{-}_2\right) = \mathcal{T}_{\hat{\vb*{x}}_{1}}\left(\delta\vb*{\rho}_{12},\delta\hat{\vb*{r}}^{-}_2\right)\\
    \left[\hat{\vb*{x}}^{-}_{2}\right] &= \mathcal{T}_{\hat{\vb*{x}}^{-}_{2}}\left(\delta\rho_1,\delta\vb*{v}_1\right)\circ\mathcal{T}_{\delta\vb*{v}_{1}}\left(\delta\vb*{\rho}_{12},\delta\hat{\vb*{r}}^{-}_2\right) = \mathcal{T}_{\hat{\vb*{x}}^{-}_{2}}\left(\delta\vb*{\rho}_{12},\delta\hat{\vb*{r}}^{-}_2\right)
\end{aligned}    
\end{equation}
Then, the continuity of the position vectors at $t_2$ is enforced by evaluating \cref{eq:o1_da_x12m_comp} in $\left(\delta\vb*{\rho}_{12},-\Delta\hat{\vb*{r}}^{-}_2\right)$ to obtain
\begin{equation}
\label{eq:o1_da_x12m_comp_eval}
\begin{aligned}
    \left[\hat{\vb*{x}}_1\right] &= \mathcal{T}_{\hat{\vb*{x}}_1}\left(\delta\vb*{\rho}_{12},\delta\hat{\vb*{r}}^{-}_2\right)\circ\left(\delta\vb*{\rho}_{12},-\Delta\hat{\vb*{r}}^{-}_2\right) = \mathcal{T}_{\hat{\vb*{x}}_1}\left(\delta\vb*{\rho}_{12}\right)\\
    \left[\hat{\vb*{x}}^{-}_2\right] &= \mathcal{T}_{\hat{\vb*{x}}^{-}_2}\left(\delta\vb*{\rho}_{12},\delta\hat{\vb*{r}}^{-}_2\right)\circ\left(\delta\vb*{\rho}_{12},-\Delta\hat{\vb*{r}}^{-}_2\right) = \mathcal{T}_{\hat{\vb*{x}}^{-}_2}\left(\delta\vb*{\rho}_{12}\right)\\
\end{aligned}    
\end{equation}
Next, the described procedure is repeated starting from the state vector at $t_3$ and propagating backward to obtain $\left[\hat{\vb*{x}}^{+}_{2}\right]$. After inverting the map of residuals and imposing the continuity in position as in \cref{eq:o1_da_delta_rr2,eq:o1_map_rho1,eq:o1_dv1,eq:o1_da_x12m_comp,eq:o1_da_x12m_comp_eval}, the following expressions are obtained
\begin{equation}
\label{eq:o1_da_x32p_comp_eval}
\begin{aligned}
    \left[\hat{\vb*{x}}^{-}_{2}\right] &= \mathcal{T}_{\hat{\vb*{x}}^{-}_{2}}\left(\delta\vb*{\rho}_{12}\right)\\
    \left[\hat{\vb*{x}}^{+}_{2}\right] &= \mathcal{T}_{\hat{\vb*{x}}^{+}_{2}}\left(\delta\vb*{\rho}_{23}\right)\\
\end{aligned}    
\end{equation}
where $\delta\vb*{\rho}_{23}=\left\{\delta\rho_2,\delta\rho_3\right\}$. At this point, the state vector at the middle epoch is continuous in position but not in velocity. The latter is enforced by computing the residuals from \cref{eq:o1_da_x32p_comp_eval} as
\begin{equation}
\begin{aligned}
    \left[\Delta\hat{\vb*{v}}_{2}\right] = \left[\hat{\vb*{v}}^{+}_{2}\right]-\left[\hat{\vb*{v}}^{-}_{2}\right] &= \mathcal{T}_{\hat{\vb*{v}}^{+}_{2}}\left(\delta\vb*{\rho}_{23}\right)-
    \mathcal{T}_{\hat{\vb*{v}}^{-}_{2}}\left(\delta\vb*{\rho}_{12}\right)\\
    &=\mathcal{T}_{\Delta\hat{\vb*{v}}_{2}}\left(\delta\vb*{\rho}\right)\\
    &= \Delta\hat{\vb*{v}}_{2}+\mathcal{T}_{\delta\hat{\vb*{v}}_{2}}\left(\delta\vb*{\rho}\right)
\end{aligned}   
\end{equation}
where $\delta\vb*{\rho} = \left\{\rho_1,\rho_2,\rho_3\right\}$. Then, the Taylor expansion $\mathcal{T}_{\delta\vb*{v}_2}\left(\delta\vb*{\rho}\right)$ is inverted to express $\delta\vb*{\rho}$ as a function of $\delta\hat{\vb*{v}}_2$, that is
\begin{equation}
\label{eq:o1_da_drho}
    \delta\vb*{\rho} = \mathcal{T}_{\delta\vb*{\rho}}\left(\delta\hat{\vb*{v}}_{2}\right)
\end{equation}
Finally, the continuity of the velocity vector at $t_2$ is enforced by evaluating \cref{eq:o1_da_drho} in $-\Delta\hat{\vb*{v}}_{2}$ to obtain the required ranges updates
\begin{equation}
\label{eq:o1_deltarho}
\Delta\vb*{\rho} = \mathcal{T}_{\delta\vb*{\rho}}\left(-\Delta\hat{\vb*{v}}_{2}\right)
\end{equation}
This correction is then plugged into the initialization of the $\rho_j$ variables, and the entire process is iterated until the correction is below a predefined threshold, i.e. $\norm{\Delta\vb*{\rho}}<\varepsilon$.

\paragraph{State estimate expansion}
\label{sec:o_iod_state_exp}

Once a solution for the nominal state is obtained, \glsentryshort{da} can be exploited to estimate the associated uncertainty due to sensor noise. 

Consider the available measurements at epochs $t_1$, $t_2$ and $t_3$ and initialize them as \glsentryshort{da} variables
\begin{equation}
\label{eq:o2_da_meas}
\begin{aligned}
    \left[\alpha_j\right] &= \bar{\alpha}_j+\beta_{\alpha_j}\delta\alpha_j\\
    \left[\delta_j\right] &= \bar{\delta}_j+\beta_{\delta_j}\delta\delta_j\\
\end{aligned}    
\end{equation}
where $\bar{\alpha}_j,\bar{\delta}_j$ are the nominal measurements and $\beta_{\alpha_j},\beta_{\delta_j}$ are the uncertainty scaling factors given by \cref{eq:da_init_ci} or \cref{eq:da_init_3s} for $j=1,2,3$. Similarly, the estimated ranges are expressed as
\begin{equation}
\label{eq:o2_da_rho}
    \left[\rho_j\right] = \rho_j+\delta\rho_j
\end{equation}
where the $\rho_j$ result from the estimation process described in \cref{sec:o_iod_range_est}. Knowing the lines of sight and the ranges, the Taylor expansions of the inertial position vectors are computed as
\begin{equation}
\label{eq:o2_da_rr}
    \left[\vb*{r}_j\right]=\mathcal{T}_{\vb*{r}_j}\left(\delta\alpha_j,\delta\delta_j,\delta \rho_j\right)  
\end{equation}
The orbital state at epoch $t_1$ is then given by
\begin{equation}
\label{eq:o2_x1}
    \left[\vb*{x}_1\right] = 
    \begin{Bmatrix}
    \left[\vb*{r}_1\right]\\
    \left[\vb*{v}_1\right]
    \end{Bmatrix}=\\
    \begin{Bmatrix}
    \mathcal{T}_{\vb*{r}_1}\left(\delta\alpha_1,\delta\delta_1,\delta\rho_1\right)\\
    \vb*{v}_{1}+\delta\vb*{v}_1
    \end{Bmatrix}=\\
    \mathcal{T}_{\vb*{x}_1}\left(\delta\alpha_1,\delta\delta_1,\delta\rho_1,\delta\vb*{v}_1\right)
\end{equation}
where $\vb*{v}_{1}$ is obtained from the range estimation phase. The state is then propagated to $t_2$ under the perturbed orbital dynamics, thus obtaining
\begin{equation}
\label{eq:o2_da_x2m}
    \left[\hat{\vb*{x}}^{-}_{2}\right]=\mathcal{T}_{\hat{\vb*{x}}^{-}_{2}}\left(\delta\alpha_1,\delta\delta_1,\delta\rho_1,\delta\vb*{v}_1\right)
\end{equation}
The position residuals at $t_2$ are then computed from \cref{eq:o2_da_x2m,eq:o2_da_rr} as
\begin{equation}
\label{eq:o2_da_delta_rr2}
    \left[\Delta\hat{\vb*{r}}_2\right] = \left[\hat{\vb*{r}}^{-}_{2}\right]-\left[\vb*{r}_{2}\right] = \Delta\hat{\vb*{r}}^{-}_2 +\mathcal{T}_{\delta\hat{\vb*{r}}^{-}_2}\left(\delta\vb*{\alpha}_{12},\delta\vb*{\delta}_{12},\delta\vb*{\rho}_{12},\delta\vb*{v}_1\right)
\end{equation}
where $\delta\vb*{\alpha}_{12}=\left\{\delta\alpha_1,\delta\alpha_2\right\}$, $\delta\vb*{\delta}_{12}=\left\{\delta\delta_1,\delta\delta_2\right\}$. An augmented map is then built as
\begin{equation}
\label{eq:dr2_map_dr}
    \begin{Bmatrix}
        \delta\vb*{\alpha}_{12}\\
        \delta\vb*{\delta}_{12}\\
        \delta\vb*{\rho}_{12}\\
        \delta\hat{\vb*{r}}^{-}_2\\
    \end{Bmatrix}=
    \begin{Bmatrix}
        \delta\vb*{\alpha}_{12}\\
        \delta\vb*{\delta}_{12}\\
        \delta\vb*{\rho}_{12}\\
        \mathcal{T}_{\delta\hat{\vb*{r}}^{-}_{2}}\left(\delta\vb*{\alpha}_{12},\delta\vb*{\delta}_{12},\delta\vb*{\rho}_{12},\delta\vb*{v}_1\right)\\
    \end{Bmatrix}
\end{equation}
and map inversion is used to obtain
\begin{equation}
\label{eq:o2_da_dv1}
    \delta\vb*{v}_1 = \mathcal{T}_{\delta\vb*{v}_1}\left(\delta\vb*{\alpha}_{12},\delta\vb*{\delta}_{12},\delta\vb*{\rho}_{12},\delta\hat{\vb*{r}}^{-}_2\right)
\end{equation}
By composing \cref{eq:o2_x1} and~\eqref{eq:o2_da_x2m} with \cref{eq:o2_da_dv1}, and then evaluating the resulting polynomials in $\left(\delta\vb*{\alpha}_{12},\delta\vb*{\delta}_{12},\delta\vb*{\rho}_{12},-\Delta\hat{\vb*{r}}^{-}_2\right)$, the following polynomial expansions are obtained
\begin{equation}
\begin{aligned}
\label{eq:o2_da_x12m_comp_eval}
    \left[\hat{\vb*{x}}_{1}\right]&=
    \mathcal{T}_{\hat{\vb*{x}}_{1}}\left(\delta\vb*{\alpha}_{12},\delta\vb*{\delta}_{12},\delta\vb*{\rho}_{12}\right)\\
    \left[\hat{\vb*{x}}^{-}_{2}\right]&=
    \mathcal{T}_{\hat{\vb*{x}}^{-}_{2}}\left(\delta\vb*{\alpha}_{12},\delta\vb*{\delta}_{12},\delta\vb*{\rho}_{12}\right)
\end{aligned}    
\end{equation}
The same procedure is then repeated starting from the orbital state at $t_3$, thus obtaining
\begin{equation}
\begin{aligned}
\label{eq:o2_da_x32p_comp_eval}
    \left[\hat{\vb*{x}}^{+}_{2}\right]&=
    \mathcal{T}_{\hat{\vb*{x}}^{+}_{2}}\left(\delta\vb*{\alpha}_{23},\delta\vb*{\delta}_{23},\delta\vb*{\rho}_{23}\right)\\
    \left[\hat{\vb*{x}}_{3}\right]&=
    \mathcal{T}_{\hat{\vb*{x}}_{3}}\left(\delta\vb*{\alpha}_{23},\delta\vb*{\delta}_{23},\delta\vb*{\rho}_{23}\right)
\end{aligned}    
\end{equation}
where $\delta\vb*{\alpha}_{23}=\left\{\delta\alpha_2,\delta\alpha_3\right\}$, $\delta\vb*{\delta}_{23}=\left\{\delta\delta_2,\delta\delta_3\right\}$, and $\delta\vb*{\rho}_{23}=\left\{\delta\rho_2,\delta\rho_3\right\}$. At this point, starting from the Taylor expansions $\left[\hat{\vb*{x}}^{+}_{2}\right]$ and $\left[\hat{\vb*{x}}^{-}_{2}\right]$, the velocity residuals are expressed as
\begin{equation}
\begin{aligned}
    \left[\Delta\hat{\vb*{v}}_{2}\right] = \left[\hat{\vb*{v}}^{+}_{2}\right]-\left[\hat{\vb*{v}}^{-}_{2}\right] &= \mathcal{T}_{\hat{\vb*{v}}^{+}_{2}}\left(\delta\vb*{\alpha}_{23},\delta\vb*{\delta}_{23},\delta\vb*{\rho}_{23}\right)-
    \mathcal{T}_{\hat{\vb*{v}}^{-}_{2}}\left(\delta\vb*{\alpha}_{12},\delta\vb*{\delta}_{12},\delta\vb*{\rho}_{12}\right)\\
    &=\mathcal{T}_{\Delta\hat{\vb*{v}}_{2}}\left(\delta\vb*{\alpha},\delta\vb*{\delta},\delta\vb*{\rho}\right)\\
    &=\Delta\hat{\vb*{v}}_{2}+\mathcal{T}_{\delta\hat{\vb*{v}}_{2}}\left(\delta\vb*{\alpha},\delta\vb*{\delta},\delta\vb*{\rho}\right)
\end{aligned}   
\end{equation}
An augmented map is then built as
\begin{equation}
\label{eq:o2_map_dv}
    \begin{Bmatrix}
        \delta\vb*{\alpha}\\
        \delta\vb*{\delta}\\
        \delta\hat{\vb*{v}}_2\\
    \end{Bmatrix}=
    \begin{Bmatrix}
        \delta\vb*{\alpha}\\
        \delta\vb*{\delta}\\
        \mathcal{T}_{\delta\hat{\vb*{v}}_2}\left(\delta\vb*{\alpha},\delta\vb*{\delta},\delta\vb*{\rho}\right)
    \end{Bmatrix}
\end{equation}
and map inversion is used to obtain
\begin{equation}
\label{eq:o2_da_drho}
    \delta\vb*{\rho} = \mathcal{T}_{\delta\vb*{\rho}}\left(\delta\vb*{\alpha},\delta\vb*{\delta},\delta\hat{\vb*{v}}_{2}\right)
\end{equation}
The dependency of $\left[\hat{\vb*{x}}_{1}\right]$ on $\delta\vb*{\rho}_{12}$ is then removed by composing \cref{eq:o2_da_x12m_comp_eval} with \cref{eq:o2_da_drho} 
\begin{equation}
\begin{aligned}
    \left[\hat{\vb*{x}}_1\right]&=
    \mathcal{T}_{\hat{\vb*{x}}_1}\left(\delta\vb*{\alpha}_{12},\delta\vb*{\delta}_{12},\delta\vb*{\rho}_{12}\right)\circ\mathcal{T}_{\delta\vb*{\rho}}\left(\delta\vb*{\alpha},\delta\vb*{\delta},\delta\hat{\vb*{v}}_2\right)\\
    &=\mathcal{T}_{\hat{\vb*{x}}_1}\left(\delta\vb*{\alpha},\delta\vb*{\delta},\delta\hat{\vb*{v}}_2\right)
\end{aligned}
\label{eq:o2_da_x12_comp}
\end{equation}
Evaluating \cref{eq:o2_da_x12_comp} in $\left(\delta\vb*{\alpha},\delta\vb*{\delta},-\Delta\hat{\vb*{v}}_2\right)$ finally yields to
\begin{equation}
\label{eq:o2_da_x12_comp_eval}
    \left[\hat{\vb*{x}}_1\right]=
    \mathcal{T}_{\hat{\vb*{x}}_1}\left(\delta\vb*{\alpha},\delta\vb*{\delta},\delta\hat{\vb*{v}}_2\right)\circ\left(\delta\vb*{\alpha},\delta\vb*{\delta},-\Delta\hat{\vb*{v}}_2\right) = \mathcal{T}_{\hat{\vb*{x}}_1}\left(\delta\vb*{\alpha},\delta\vb*{\delta}\right)
\end{equation}
which is the Taylor expansion of the orbital solution with respect to uncertainties in the available measurements.

\section{Numerical simulations}\label{sec:num_simu}
This section describes the performance of the proposed method obtained through numerical simulations. For all simulations, an expansion order of four was used. To control the \glsentryshort{ads} algorithm, the tolerances on the components of the state vector (the output of the function wrapped by the \glsentryshort{ads} algorithm) were set to \SI{10}{\meter} and \SI{1}{\milli\meter\per\second} in position and velocity, respectively. However, no splits were observed in the analyzed scenarios. A discussion on the performance of the \glsentryshort{ads} algorithm in the context of \glsentryshort{iod} is provided in~\cite{Pirovano2020e} for optical measurements under the assumption of Keplerian dynamics. The conclusions presented in that paper are considered representative of the solution that would be obtained with the same input measurements by considering $J_2$-perturbed dynamics. All the simulations were run on an Intel i7-8565U CPU @1.80 GHz and 16 GB of RAM. The algorithms were implemented in Java and interface with the CNES library \glsentryfullinv{pace} to perform all \glsentryshort{da} operations.

Following the scheme described in~\cite{Losacco2023}, simulations were performed by considering a subset of the NORAD \glsentryfull{leo} population. The analyses were carried out by downloading the latest \glsentryfull{tle}, considering a one-day propagation window in high-fidelity dynamics (including the Earth gravitational potential up to order and degree 8, the third body effect of the Sun and the Moon, the atmospheric drag, and the solar radiation pressure), and then generating the measurements produced by different observers. In these analyses, two observers were considered: a Doppler-only radar and an optical telescope. The Doppler-only radar has a bistatic configuration, where the receiver has latitude 44$^{\circ}$ 4$^{'}$ 17$^{''}$ North, longitude 5$^{\circ}$ 32$^{'}$ 4$^{''}$ East and altitude 180~m, while the transmitter has latitude 47$^{\circ}$ 20$^{'}$ 53$^{''}$ North, longitude 5$^{\circ}$ 30$^{'}$ 54$^{''}$ East and altitude 180~m. The transmission site is assumed to be capable of covering the azimuth band from 90~deg to 270~deg and the elevation band from 20~deg to 40~deg, whereas the field-of-view of the receiver is assumed to be infinite, i.e., an object is observed whenever it is detected by the transmitter. Instead, the optical sensor was co-located with the radar transmitter. To increase the number of optical detections, no specific pointing direction or field-of-view were selected, i.e., the optical sensor was assumed capable of covering all passages detected by the radar. This scenario is obviously unrealistic but allows us to investigate a wider selection of cases. Approximately 2,000 passages were generated. For each passage, ten different measurement noise levels were considered. These noise levels are indicated with the symbols $k_i^{\sigma}$, with $i=1,\ldots,10$. For the Doppler radar, they range from $k_1^{\sigma}=\left(0.01\ \textrm{deg};\ 0.01\ \textrm{deg};\ 0.1\ \textrm{m/s}\right)$ to $k_{10}^{\sigma} = 10 k_1^{\sigma}$, with a step $\Delta k^{\sigma}=k_{i+1}^{\sigma}-k_i^{\sigma}=k_1^{\sigma}$, with $i=1,\ldots,9$, where each triplet indicates the noise standard deviations in azimuth, elevation, and range rate, respectively. For the optical telescope, $k_1^{\sigma}=\left(0.1\ \textrm{arcsec};\ 0.1\ \textrm{arcsec}\right)$ while $k_{10}^{\sigma} = 10 k_1^{\sigma}$, with $\Delta k^{\sigma}=k_1^{\sigma}$, where each couple indicates the noise standard deviations in right ascension and declination. A total of 20,000 passages were generated. For each passage, the proposed methods were executed by considering two different orbital dynamics: Keplerian and $J_2$-perturbed dynamics~\cite{Armellin2018h}. The aim was to investigate the variation in the performance of the proposed methods as a function of the selected dynamics, arc length, and noise level. Two indices were considered: $\varepsilon_{\vb*{x}}$ and $f_{\vb*{x}}$. The first index is the nondimensional error between the estimated and true states, defined as
\begin{equation}
    \varepsilon_{\vb*{x}} = \norm{\dfrac{\hat{\vb*{x}}-\vb*{x}}{\vb*{\gamma}}}_2
    \label{eq:nd_err_def}
\end{equation}
where $\hat{\vb*{x}}$ is the computed estimate, $\vb*{x}$ is the true state, and $\vb*{\gamma}$ is a six-dimensional vector of scaling coefficients used to normalize the error vector components. In this work, $\vb*{\gamma}=\{R_E,R_E,R_E,v_c,v_c,v_c\}^T$ with $R_E$ the equatorial radius of the Earth, $v_c=\sqrt{\mu/R_E}$ the orbital velocity on a circular orbit at $R_E$, and $\mu$ the Earth standard gravitational parameter. Vector subtraction and division in \cref{eq:nd_err_def} are performed element-wise such that the argument of $\norm{\cdot}_2$ is a nondimensional error vector whose Euclidean norm is defined as $\varepsilon_{\vb*{x}}$. The second index $f_{\vb*{x}}$ is the fraction of the estimated bounds that correctly include the true state and is computed as follows: given the polynomial expansion of the \glsentryshort{iod} solution $[\hat{\vb*{x}}]$, \glsentryshort{da} routines are used to estimate the lower and upper bounds $\vb*{x}^{lb},\vb*{x}^{ub}$ within which the former expansion satisfies the accuracy imposed by the \glsentryshort{ads} algorithm. As such, any state $\tilde{\vb*{x}}$ that falls within these bounds, i.e. $\tilde{\vb*{x}}_i\in[\vb*{x}^{lb}_i,\vb*{x}^{ub}_i]\ \forall i\in[1,6]$, can be accurately mapped to the deviation $\delta\tilde{\vb*{y}}$ with respect to the nominal measurement vector $\vb*{y}$ that corresponds to the displaced state $\tilde{\vb*{x}}$. Since the truncation error within these bounds is controlled by the \glsentryshort{ads} algorithm, it is of interest to quantify the ratio between the number of tuples $[\vb*{x}^{lb}_i,\vb*{x}^{ub}_i]$ that include the true solution $\vb*{x}_i$ and the total number of state components, with the last equal to the dimension of $\hat{\vb*{x}}$. For each component of the state a check is thus performed, and a vector of binary values is built by assigning ``1'' to the components for which the true state falls within the estimated bounds and ``0'' to the ones for which the true state lies outside. The index $f_{\vb*{x}}$ is then taken as the arithmetic mean of these values such that $f_{\vb*{x}}\in[0,1]$.

\begin{table*}[ht!]
    \centering
    \small
    \sisetup{round-mode=places,round-precision=4,scientific-notation=true}
    \begin{tabular}{c c c c c c}
    \hline\hline
    & & {$\Delta t_{obs}<0.03T$} & {$\Delta t_{obs}<0.06T$} & {$\Delta t_{obs}<0.09T$} & {$\Delta t_{obs}<0.12T$}\\  
    \cline{2-6}   
		\multirow{5}{*}{\shortstack[c]{$\varepsilon_{\vb*{x}}$\\ (-)}}&	$k^{\sigma}_{2}$&	\num[round-mode=figures,round-precision=5]{0.0009573711}&	\num[round-mode=figures,round-precision=5]{0.0006934578}&	\num[round-mode=figures,round-precision=5]{0.0006848074}&	\num[round-mode=figures,round-precision=5]{0.0006838096}\\
&	$k^{\sigma}_{4}$&	\num[round-mode=figures,round-precision=5]{0.0017752820}&	\num[round-mode=figures,round-precision=5]{0.0011502962}&	\num[round-mode=figures,round-precision=5]{0.0011218621}& \num[round-mode=figures,round-precision=5]{0.0011175349}\\
&	$k^{\sigma}_{6}$&	\num[round-mode=figures,round-precision=5]{0.0022476386}&	\num[round-mode=figures,round-precision=5]{0.0014172628}&	\num[round-mode=figures,round-precision=5]{0.0013774026}& \num[round-mode=figures,round-precision=5]{0.0013711415}\\
&	$k^{\sigma}_{8}$&	\num[round-mode=figures,round-precision=5]{0.0029636790}&	\num[round-mode=figures,round-precision=5]{0.0018204672}&	\num[round-mode=figures,round-precision=5]{0.0017630278}&	\num[round-mode=figures,round-precision=5]{0.0017539317}\\
&	$k^{\sigma}_{10}$&	\num[round-mode=figures,round-precision=5]{0.0038940036}&	\num[round-mode=figures,round-precision=5]{0.0023384933}&	\num[round-mode=figures,round-precision=5]{0.0022588784}&	\num[round-mode=figures,round-precision=5]{0.0022459067}\\  
    \hline\hline
    \end{tabular}
    \vspace{0.2cm}
    \caption{State errors under Keplerian dynamics  (optical sensor, raw data).}
    \label{tab:Results_err_kep_5s_opt_raw}
\end{table*}

\begin{table*}[ht!]
    \centering
    \small
    \sisetup{round-mode=places,round-precision=4,scientific-notation=true}
    \begin{tabular}{c c c c c c}
    \hline\hline
    & & {$\Delta t_{obs}<0.03T$} & {$\Delta t_{obs}<0.06T$} & {$\Delta t_{obs}<0.09T$} & {$\Delta t_{obs}<0.12T$}\\ 
    \cline{2-6}
    		\multirow{5}{*}{\shortstack[c]{$\varepsilon_{\vb*{x}}$\\ (-)}}&	$k^{\sigma}_{2}$&	\num[round-mode=figures,round-precision=5]{0.0009514352}&	\num[round-mode=figures,round-precision=5]{0.0005840501}&	\num[round-mode=figures,round-precision=5]{0.0005662452}& \num[round-mode=figures,round-precision=5]{0.0005632980}\\
    		&	$k^{\sigma}_{4}$&	\num[round-mode=figures,round-precision=5]{0.0018015913}&	\num[round-mode=figures,round-precision=5]{0.0010626498}&	\num[round-mode=figures,round-precision=5]{0.0010240768}&	\num[round-mode=figures,round-precision=5]{0.0010177030}\\
    		&	$k^{\sigma}_{6}$&	\num[round-mode=figures,round-precision=5]{0.0022056888}&	\num[round-mode=figures,round-precision=5]{0.0012992008}&	\num[round-mode=figures,round-precision=5]{0.0012506895}& \num[round-mode=figures,round-precision=5]{0.0012426079}\\
    		&	$k^{\sigma}_{8}$&	\num[round-mode=figures,round-precision=5]{0.0029387165}&	\num[round-mode=figures,round-precision=5]{0.0017163353}&	\num[round-mode=figures,round-precision=5]{0.0016498968}&	\num[round-mode=figures,round-precision=5]{0.0016387864}\\
&	$k^{\sigma}_{10}$&	\num[round-mode=figures,round-precision=5]{0.0038859507}&	\num[round-mode=figures,round-precision=5]{0.0022528932}&	\num[round-mode=figures,round-precision=5]{0.0021630647}&	\num[round-mode=figures,round-precision=5]{0.0021481306}\\
    \hline\hline
    \end{tabular}
    \vspace{0.2cm}
    \caption{State errors under $J_2$ dynamics (optical sensor, raw data).}
    \label{tab:Results_err_j2_5s_opt_raw}
\end{table*}

\Cref{tab:Results_err_kep_5s_opt_raw} shows the performance of the proposed method in terms of $\varepsilon_{\vb*{x}}$ when considering Keplerian dynamics on raw data collected by the optical sensor. For conciseness, only five noise levels corresponding to $k_{2j}^{\sigma}$ for $j=1,\ldots,5$ are reported in the subsequent tables. The results are consistent across different values of $k_i^{\sigma}$ and no additional conclusions can be drawn from the omitted results. All passages were sorted according to their noise level $k_i^{\sigma}$ and observed arc length expressed as a fraction of the orbital period $\Delta t_{obs}$. The $\varepsilon_{\vb*{x}}$ index was computed for all passages falling into a specific $\left(k_i^{\sigma};\Delta t_{obs}\right)$ slot and the values reported in the table are the averages per slot. First, the trend of the estimation error is analyzed as a function of the noise level. As expected, this error increases as the measurement accuracy decreases. Conversely, for a fixed noise level and increasing arc duration, the error decreased progressively. Given the relatively short duration of the passages (\glsentryshort{leo} objects observed during a single pass), this trend is also expected, as measurements spread on a longer arc allow the solver to better capture the curvature of the underlying orbit. \Cref{tab:Results_err_j2_5s_opt_raw} shows instead the results obtained for the same test cases when the $J_2$-perturbed dynamics is considered. If on one side the trends remain the same, comparing \cref{tab:Results_err_kep_5s_opt_raw,tab:Results_err_j2_5s_opt_raw} slot by slot shows that the introduction of the $J_2$ perturbation systematically improves the accuracy of the obtained solution.

\begin{table*}[ht!]
    \centering
    \small
    \sisetup{round-mode=places,round-precision=4,group-digits=integer}
    \begin{tabular}{c c c c c c}
    \hline\hline
    & & {$\Delta t_{obs}<0.03T$} & {$\Delta t_{obs}<0.06T$} & {$\Delta t_{obs}<0.09T$} & {$\Delta t_{obs}<0.12T$}\\  
    \cline{2-6}   
		\multirow{5}{*}{\shortstack[c]{$f_{\vb*{x}}$\\ (-)}}&	$k^{\sigma}_{2}$&	\num[round-mode=figures,round-precision=5]{0.9369240547}&	\num[round-mode=figures,round-precision=5]{0.7185850053}&	\num[round-mode=figures,round-precision=5]{0.6891414141}&	\num[round-mode=figures,round-precision=5]{0.6843776107}\\
&	$k^{\sigma}_{4}$&	\num[round-mode=figures,round-precision=5]{0.9863100923}&	\num[round-mode=figures,round-precision=5]{0.8488208377}&	\num[round-mode=figures,round-precision=5]{0.8176767677}&	\num[round-mode=figures,round-precision=5]{0.8123642439}\\
&	$k^{\sigma}_{6}$&	\num[round-mode=figures,round-precision=5]{0.9955470738}&	\num[round-mode=figures,round-precision=5]{0.9139387540}&	\num[round-mode=figures,round-precision=5]{0.8860269360}&	\num[round-mode=figures,round-precision=5]{0.8807017544}\\
&	$k^{\sigma}_{8}$&	\num[round-mode=figures,round-precision=5]{0.9982456140}&	\num[round-mode=figures,round-precision=5]{0.9439738839}&	\num[round-mode=figures,round-precision=5]{0.9237974684}& \num[round-mode=figures,round-precision=5]{0.9186767169}\\
&	$k^{\sigma}_{10}$&	\num[round-mode=figures,round-precision=5]{0.9992010227}&	\num[round-mode=figures,round-precision=5]{0.9633156966}&	\num[round-mode=figures,round-precision=5]{0.9467780027}& \num[round-mode=figures,round-precision=5]{0.9419889503}\\
    \hline\hline
    \end{tabular}
    \vspace{0.2cm}
    \caption{Bound success under Keplerian dynamics assumption (optical sensor, raw data).}
    \label{tab:Results_bd_success_kep_5s_opt_raw}
\end{table*}

\begin{table*}[ht!]
    \centering
    \small
    \sisetup{round-mode=places,round-precision=4,group-digits=integer}
    \begin{tabular}{c c c c c c}
    \hline\hline
    & & {$\Delta t_{obs}<0.03T$} & {$\Delta t_{obs}<0.06T$} & {$\Delta t_{obs}<0.09T$} & {$\Delta t_{obs}<0.12T$}\\ 
    \cline{2-6}
		\multirow{5}{*}{\shortstack[c]{$f_{\vb*{x}}$\\ (-)}}&	$k^{\sigma}_{2}$&	\num[round-mode=figures,round-precision=5]{0.9869717191}&	\num[round-mode=figures,round-precision=5]{0.8855156635}&	\num[round-mode=figures,round-precision=5]{0.8675925926}&	\num[round-mode=figures,round-precision=5]{0.8649958229}\\
&	$k^{\sigma}_{4}$&	\num[round-mode=figures,round-precision=5]{0.9968163005}&	\num[round-mode=figures,round-precision=5]{0.9418338613}&	\num[round-mode=figures,round-precision=5]{0.9276936027}&	\num[round-mode=figures,round-precision=5]{0.9256474520}\\
&	$k^{\sigma}_{6}$&	\num[round-mode=figures,round-precision=5]{0.9985687023}&	\num[round-mode=figures,round-precision=5]{0.9638331573}&	\num[round-mode=figures,round-precision=5]{0.9529461279}&	\num[round-mode=figures,round-precision=5]{0.9508771930}\\
&	$k^{\sigma}_{8}$&	\num[round-mode=figures,round-precision=5]{0.9993620415}&	\num[round-mode=figures,round-precision=5]{0.9745014999}&	\num[round-mode=figures,round-precision=5]{0.9659915612}&	\num[round-mode=figures,round-precision=5]{0.9639028476}\\
&	$k^{\sigma}_{10}$&	\num[round-mode=figures,round-precision=5]{0.9996804091}&	\num[round-mode=figures,round-precision=5]{0.9826278660}&	\num[round-mode=figures,round-precision=5]{0.9747807018}&	\num[round-mode=figures,round-precision=5]{0.9726268207}\\
    \hline\hline
    \end{tabular}
    \vspace{0.2cm}
    \caption{Bound success under $J_2$ dynamics assumption (optical sensor, raw data).}
    \label{tab:Results_bd_success_j2_5s_opt_raw}
\end{table*}

The advantages of a more refined dynamics can easily be observed by investigating the accuracy of the estimated bounds. \Cref{tab:Results_bd_success_kep_5s_opt_raw} shows the trend of the average $f_{\vb*{x}}$ parameter as a function of noise level and arc length. Let first set the noise level to $k_2^{\sigma}$ and consider increasingly longer observation arcs. As can be seen, the percentage of success progressively decreases from 93.69$\%$ to 68.44$\%$. That is, the longer the arc, the less accurate the estimated bounds are. This trend is the opposite of that identified for $\varepsilon_{\vb*{x}}$, for which an increase in the observed arc length is beneficial. This is because as the passage duration increases, the solution becomes increasingly accurate, but the estimated bounds shrink excessively around the nominal solution up to a point where they often no longer include the real state. This undesired behavior can be explained by considering all the actors coming into play: noise level, arc length, and dynamics. As the arc length increases, the Keplerian assumption becomes less accurate. Although longer arcs allow to better estimate the orbit, the improvements are insufficient to keep up with the shrinking of the estimated bounds, which are therefore unreliable. This situation becomes less critical with the increase in the noise level: the bounds are in this case inflated by the larger uncertainty in the measurements and the mismatch in the dynamics has almost no impact on their accuracy. This is evidenced by the line for $k_{10}^{\sigma}$ where $\varepsilon_{\vb*{x}}$ increases from 99.92$\%$ to 94.20$\%$. The introduction of a higher fidelity dynamics partially mitigates this trend and provides a more accurate solution as the considered arc length increases, as shown in \cref{tab:Results_bd_success_j2_5s_opt_raw}. In this case, the advantage of considering the $J_2$ perturbation progressively increases for longer passages, with an average success rate always above 86$\%$.


\begin{table*}[ht!]
    \centering
    \small
    \sisetup{round-mode=places,round-precision=4,scientific-notation=true}
    \begin{tabular}{c c c c c c}
    \hline\hline
    & & {$\Delta t_{obs}<0.03T$} & {$\Delta t_{obs}<0.06T$} & {$\Delta t_{obs}<0.09T$} & {$\Delta t_{obs}<0.12T$}\\  
    \cline{2-6}  
		\multirow{5}{*}{\shortstack[c]{$\varepsilon_{\vb*{x}}$\\ (-)}}&	$k^{\sigma}_{2}$&	\num[round-mode=figures,round-precision=5]{0.0008663058}&	\num[round-mode=figures,round-precision=5]{0.0006423378}&	\num[round-mode=figures,round-precision=5]{0.0006359618}&	\num[round-mode=figures,round-precision=5]{0.0006353212}\\
&	$k^{\sigma}_{4}$&	\num[round-mode=figures,round-precision=5]{0.0013873965}&	\num[round-mode=figures,round-precision=5]{0.0009338005}&	\num[round-mode=figures,round-precision=5]{0.0009148085}&	\num[round-mode=figures,round-precision=5]{0.0009120919}\\
&	$k^{\sigma}_{6}$&	\num[round-mode=figures,round-precision=5]{0.0017111763}&	\num[round-mode=figures,round-precision=5]{0.0011186654}&	\num[round-mode=figures,round-precision=5]{0.0010917110}&	\num[round-mode=figures,round-precision=5]{0.0010876276}\\
&	$k^{\sigma}_{8}$&	\num[round-mode=figures,round-precision=5]{0.0022264964}&	\num[round-mode=figures,round-precision=5]{0.0014056306}&	\num[round-mode=figures,round-precision=5]{0.0013660472}&	\num[round-mode=figures,round-precision=5]{0.0013598540}\\
&	$k^{\sigma}_{10}$&	\num[round-mode=figures,round-precision=5]{0.0027562235}&	\num[round-mode=figures,round-precision=5]{0.0017006443}&	\num[round-mode=figures,round-precision=5]{0.0016485912}&	\num[round-mode=figures,round-precision=5]{0.0016402863}\\  
    \hline\hline
    \end{tabular}
    \vspace{0.2cm}
    \caption{State errors under Keplerian dynamics assumption (optical, regressed data).}
    \label{tab:Results_err_kep_5s_opt_regr}
\end{table*}

\begin{table*}[ht!]
    \centering
    \small
    \sisetup{round-mode=places,round-precision=4,scientific-notation=true}
    \begin{tabular}{c c c c c c}
    \hline\hline
    & & {$\Delta t_{obs}<0.03T$} & {$\Delta t_{obs}<0.06T$} & {$\Delta t_{obs}<0.09T$} & {$\Delta t_{obs}<0.12T$}\\ 
    \cline{2-6}
 		\multirow{5}{*}{\shortstack[c]{$\varepsilon_{\vb*{x}}$\\ (-)}}&	$k^{\sigma}_{2}$&	\num[round-mode=figures,round-precision=5]{0.0008238483}&	\num[round-mode=figures,round-precision=5]{0.0005105338}&	\num[round-mode=figures,round-precision=5]{0.0004959390}&	\num[round-mode=figures,round-precision=5]{0.0004935225}\\
&	$k^{\sigma}_{4}$&	\num[round-mode=figures,round-precision=5]{0.0013577777}&	\num[round-mode=figures,round-precision=5]{0.0008097810}&	\num[round-mode=figures,round-precision=5]{0.0007821885}&	\num[round-mode=figures,round-precision=5]{0.0007776269}\\
&	$k^{\sigma}_{6}$&	\num[round-mode=figures,round-precision=5]{0.0016523542}&	\num[round-mode=figures,round-precision=5]{0.0009798016}&	\num[round-mode=figures,round-precision=5]{0.0009449327}&	\num[round-mode=figures,round-precision=5]{0.0009391522}\\
&	$k^{\sigma}_{8}$&	\num[round-mode=figures,round-precision=5]{0.0022055253}&	\num[round-mode=figures,round-precision=5]{0.0012929746}&	\num[round-mode=figures,round-precision=5]{0.0012445924}&	\num[round-mode=figures,round-precision=5]{0.0012365501}\\
&	$k^{\sigma}_{10}$&	\num[round-mode=figures,round-precision=5]{0.0027479534}&	\num[round-mode=figures,round-precision=5]{0.0015950781}&	\num[round-mode=figures,round-precision=5]{0.0015331823}&	\num[round-mode=figures,round-precision=5]{0.0015229279}\\   
    \hline\hline
    \end{tabular}
    \vspace{0.2cm}
    \caption{State errors under $J_2$ dynamics assumption (optical, regressed data)}
    \label{tab:Results_err_j2_5s_opt_regr}
\end{table*}

\begin{table*}[ht!]
    \centering
    \small
    \sisetup{round-mode=places,round-precision=4,group-digits=integer}
    \begin{tabular}{c c c c c c}
    \hline\hline
    & & {$\Delta t_{obs}<0.03T$} & {$\Delta t_{obs}<0.06T$} & {$\Delta t_{obs}<0.09T$} & {$\Delta t_{obs}<0.12T$}\\  
    \cline{2-6}  
		\multirow{5}{*}{\shortstack[c]{$f_{\vb*{x}}$\\ (-)}}&	$k^{\sigma}_{2}$&	\num[round-mode=figures,round-precision=5]{0.8326990168}&	\num[round-mode=figures,round-precision=5]{0.5728837373}&	\num[round-mode=figures,round-precision=5]{0.5489751344}&	\num[round-mode=figures,round-precision=5]{0.5449391362}\\
&	$k^{\sigma}_{4}$&	\num[round-mode=figures,round-precision=5]{0.9373015873}&	\num[round-mode=figures,round-precision=5]{0.7157793007}&	\num[round-mode=figures,round-precision=5]{0.6862497899}&	\num[round-mode=figures,round-precision=5]{0.6813480147}\\
&	$k^{\sigma}_{6}$&	\num[round-mode=figures,round-precision=5]{0.9717190976}&	\num[round-mode=figures,round-precision=5]{0.7877748461}&	\num[round-mode=figures,round-precision=5]{0.7562678782}&	\num[round-mode=figures,round-precision=5]{0.7510020040}\\
&	$k^{\sigma}_{8}$&	\num[round-mode=figures,round-precision=5]{0.9864563416}&	\num[round-mode=figures,round-precision=5]{0.8436344427}&	\num[round-mode=figures,round-precision=5]{0.8114367526}&	\num[round-mode=figures,round-precision=5]{0.8060013373}\\
&	$k^{\sigma}_{10}$&	\num[round-mode=figures,round-precision=5]{0.9866730893}&	\num[round-mode=figures,round-precision=5]{0.8748894783}&	\num[round-mode=figures,round-precision=5]{0.8429731101}&	\num[round-mode=figures,round-precision=5]{0.8374454515}\\    
    \hline\hline
    \end{tabular}
    \vspace{0.2 cm}
    \caption{Bound success under Keplerian dynamics assumption (optical, regressed data).}
    \label{tab:Results_bd_success_kep_5s_opt_regr}
\end{table*}

\begin{table*}[ht!]
    \centering
    \small
    \sisetup{round-mode=places,round-precision=4,group-digits=integer}
    \begin{tabular}{c c c c c c}
    \hline\hline
    & & {$\Delta t_{obs}<0.03T$} & {$\Delta t_{obs}<0.06T$} & {$\Delta t_{obs}<0.09T$} & {$\Delta t_{obs}<0.12T$}\\ 
    \cline{2-6}
		\multirow{5}{*}{\shortstack[c]{$f_{\vb*{x}}$\\ (-)}}&	$k^{\sigma}_{2}$&	\num[round-mode=figures,round-precision=5]{0.9659054868}&	\num[round-mode=figures,round-precision=5]{0.8078679312}&	\num[round-mode=figures,round-precision=5]{0.7881384409}&	\num[round-mode=figures,round-precision=5]{0.7837251959}\\
&	$k^{\sigma}_{4}$&	\num[round-mode=figures,round-precision=5]{0.9876190476}&	\num[round-mode=figures,round-precision=5]{0.8871903005}&	\num[round-mode=figures,round-precision=5]{0.8690536225}&	\num[round-mode=figures,round-precision=5]{0.8659492826}\\
&	$k^{\sigma}_{6}$&	\num[round-mode=figures,round-precision=5]{0.9936447410}&	\num[round-mode=figures,round-precision=5]{0.9215479332}&	\num[round-mode=figures,round-precision=5]{0.9049301699}&	\num[round-mode=figures,round-precision=5]{0.9019706079}\\
&	$k^{\sigma}_{8}$&	\num[round-mode=figures,round-precision=5]{0.9961759082}&	\num[round-mode=figures,round-precision=5]{0.9407466103}&	\num[round-mode=figures,round-precision=5]{0.9250463197}&	\num[round-mode=figures,round-precision=5]{0.9225175527}\\
&	$k^{\sigma}_{10}$&	\num[round-mode=figures,round-precision=5]{0.9963070006}&	\num[round-mode=figures,round-precision=5]{0.9529619805}&	\num[round-mode=figures,round-precision=5]{0.9389480805}&	\num[round-mode=figures,round-precision=5]{0.9366398120}\\    
    \hline\hline
    \end{tabular}
    \vspace{0.2cm}
    \caption{Bound success under $J_2$ dynamics assumption (optical, regressed data).}
    \label{tab:Results_bd_success_j2_5s_opt_regr}
\end{table*}

\Crefrange{tab:Results_err_kep_5s_opt_regr}{tab:Results_bd_success_j2_5s_opt_regr} show the trends for the $\varepsilon_{\vb*{x}}$ and $f_{\vb*{x}}$ indices considering both Keplerian and $J_2$ perturbed dynamics when processing regressed measurements. These trends closely resemble those found with raw data. However, the introduction of measurement regression has two major effects. The first is a general improvement in the solution accuracy $\varepsilon_{\vb*{x}}$, which can be seen by comparing \cref{tab:Results_err_kep_5s_opt_regr} with \cref{tab:Results_err_kep_5s_opt_raw} and \cref{tab:Results_err_j2_5s_opt_regr} with~\cref{tab:Results_err_j2_5s_opt_raw}. Measurement regression allows the solver to mitigate the effect of measurement noise, thus preventing the algorithm from underperforming due to possible strong local realizations of measurement noise in the three observation epochs exploited for \glsentryshort{iod}. The second trend is instead a general worsening of the accuracy on the bounds, regardless of the noise level and arc length. This behavior is also expected since the regression tends to shrink the uncertainty region around the fitted measurements. This has the undesired effect of exacerbating the impact of the orbital dynamics mismatch, which is only partially mitigated by the introduction of $J_2$ perturbations.


\begin{table*}[ht!]
    \centering
    \small
    \sisetup{round-mode=places,round-precision=4,scientific-notation=true}
    \begin{tabular}{c c c c c c}
    \hline\hline
    & & {$\Delta t_{obs}<0.03T$} & {$\Delta t_{obs}<0.06T$} & {$\Delta t_{obs}<0.09T$} & {$\Delta t_{obs}<0.12T$}\\  
    \cline{2-6}  
    		\multirow{5}{*}{\shortstack[c]{$\varepsilon_{\vb*{x}}$\\ (-)}}&	$k^{\sigma}_{2}$&	\num[round-mode=figures,round-precision=5]{0.0030774208}&	\num[round-mode=figures,round-precision=5]{0.0020678971}&	\num[round-mode=figures,round-precision=5]{0.0020047396}&	\num[round-mode=figures,round-precision=5]{0.0019947077}\\
&	$k^{\sigma}_{4}$&	\num[round-mode=figures,round-precision=5]{0.0061967624}&	\num[round-mode=figures,round-precision=5]{0.0040612199}&	\num[round-mode=figures,round-precision=5]{0.0039366980}&	\num[round-mode=figures,round-precision=5]{0.0039145132}\\
&	$k^{\sigma}_{6}$&	\num[round-mode=figures,round-precision=5]{0.0080058770}&	\num[round-mode=figures,round-precision=5]{0.0053868897}&	\num[round-mode=figures,round-precision=5]{0.0052270242}&	\num[round-mode=figures,round-precision=5]{0.0052015250}\\
&	$k^{\sigma}_{8}$&	\num[round-mode=figures,round-precision=5]{0.0118420589}&	\num[round-mode=figures,round-precision=5]{0.0077992152}&	\num[round-mode=figures,round-precision=5]{0.0075570141}&	\num[round-mode=figures,round-precision=5]{0.0075171541}\\
&	$k^{\sigma}_{10}$&	\num[round-mode=figures,round-precision=5]{0.0126011311}&	\num[round-mode=figures,round-precision=5]{0.0084968113}&	\num[round-mode=figures,round-precision=5]{0.0082419203}&	\num[round-mode=figures,round-precision=5]{0.0082036039}\\   
    \hline\hline
    \end{tabular}
    \vspace{0.2cm}
    \caption{State errors under Keplerian dynamics assumption (Doppler radar, raw data).}
    \label{tab:Results_err_kep_5s_dr_raw}
\end{table*}

\begin{table*}[ht!]
    \centering
    \small
    \sisetup{round-mode=places,round-precision=4,scientific-notation=true}
    \begin{tabular}{c c c c c c}
    \hline\hline
    & & {$\Delta t_{obs}<0.03T$} & {$\Delta t_{obs}<0.06T$} & {$\Delta t_{obs}<0.09T$} & {$\Delta t_{obs}<0.12T$}\\ 
    \cline{2-6}
    		\multirow{5}{*}{\shortstack[c]{$\varepsilon_{\vb*{x}}$\\ (-)}}&	$k^{\sigma}_{2}$&	\num[round-mode=figures,round-precision=5]{0.0022860860}&	\num[round-mode=figures,round-precision=5]{0.0016191055}&	\num[round-mode=figures,round-precision=5]{0.0015735769}&	\num[round-mode=figures,round-precision=5]{0.0015662312}\\
&	$k^{\sigma}_{4}$&	\num[round-mode=figures,round-precision=5]{0.0044077222}&	\num[round-mode=figures,round-precision=5]{0.0030635788}&	\num[round-mode=figures,round-precision=5]{0.0029812375}&	\num[round-mode=figures,round-precision=5]{0.0029661487}\\
&	$k^{\sigma}_{6}$&	\num[round-mode=figures,round-precision=5]{0.0063448399}&	\num[round-mode=figures,round-precision=5]{0.0044627474}&	\num[round-mode=figures,round-precision=5]{0.0043427158}&	\num[round-mode=figures,round-precision=5]{0.0043239016}\\
&	$k^{\sigma}_{8}$&	\num[round-mode=figures,round-precision=5]{0.0085574960}&	\num[round-mode=figures,round-precision=5]{0.0060008579}&	\num[round-mode=figures,round-precision=5]{0.0058361954}&	\num[round-mode=figures,round-precision=5]{0.0058093356}\\
&	$k^{\sigma}_{10}$&	\num[round-mode=figures,round-precision=5]{0.0100532802}&	\num[round-mode=figures,round-precision=5]{0.0071071981}&	\num[round-mode=figures,round-precision=5]{0.0069135864}&	\num[round-mode=figures,round-precision=5]{0.0068856254}\\
    \hline\hline
    \end{tabular}
    \vspace{0.2cm}
    \caption{State errors under $J_2$ dynamics assumption (Doppler radar, raw data)}
    \label{tab:Results_err_j2_5s_dr_raw}
\end{table*}

\begin{table*}[ht!]
    \centering
    \small
    \sisetup{round-mode=places,round-precision=4,group-digits=integer}
    \begin{tabular}{c c c c c c}
    \hline\hline
    & & {$\Delta t_{obs}<0.03T$} & {$\Delta t_{obs}<0.06T$} & {$\Delta t_{obs}<0.09T$} & {$\Delta t_{obs}<0.12T$}\\  
    \cline{2-6}  
    		\multirow{5}{*}{\shortstack[c]{$f_{\vb*{x}}$\\ (-)}}&	$k^{\sigma}_{2}$&	\num[round-mode=figures,round-precision=5]{0.9983912484}&	\num[round-mode=figures,round-precision=5]{0.9990227434}&	\num[round-mode=figures,round-precision=5]{0.9990651028}&	\num[round-mode=figures,round-precision=5]{0.9990721997}\\
&	$k^{\sigma}_{4}$&	\num[round-mode=figures,round-precision=5]{0.9974235105}&	\num[round-mode=figures,round-precision=5]{0.9985732121}&	\num[round-mode=figures,round-precision=5]{0.9986352781}&	\num[round-mode=figures,round-precision=5]{0.9986456746}\\
&	$k^{\sigma}_{6}$&	\num[round-mode=figures,round-precision=5]{0.9980449658}&	\num[round-mode=figures,round-precision=5]{0.9987394201}&	\num[round-mode=figures,round-precision=5]{0.9987086777}&	\num[round-mode=figures,round-precision=5]{0.9987186058}\\
&	$k^{\sigma}_{8}$&	\num[round-mode=figures,round-precision=5]{0.9958291625}&	\num[round-mode=figures,round-precision=5]{0.9976333515}&	\num[round-mode=figures,round-precision=5]{0.9977371628}&	\num[round-mode=figures,round-precision=5]{0.9977547496}\\
&	$k^{\sigma}_{10}$&	\num[round-mode=figures,round-precision=5]{0.9962887989}&	\num[round-mode=figures,round-precision=5]{0.9977017834}&	\num[round-mode=figures,round-precision=5]{0.9978035495}&	\num[round-mode=figures,round-precision=5]{0.9978207810}\\ 
    \hline\hline
    \end{tabular}
    \vspace{0.2cm}
    \caption{Bound success under Keplerian dynamics assumption (Doppler radar, raw data).}
    \label{tab:Results_bd_success_kep_5s_dr_raw}
\end{table*}

\begin{table*}[ht!]
    \centering
    \small
    \sisetup{round-mode=places,round-precision=4,group-digits=integer}
    \begin{tabular}{c c c c c c}
    \hline\hline
    & & {$\Delta t_{obs}<0.03T$} & {$\Delta t_{obs}<0.06T$} & {$\Delta t_{obs}<0.09T$} & {$\Delta t_{obs}<0.12T$}\\ 
    \cline{2-6}
    		\multirow{5}{*}{\shortstack[c]{$f_{\vb*{x}}$\\ (-)}}&	$k^{\sigma}_{2}$&	\num[round-mode=figures,round-precision=5]{0.9995173745}&	\num[round-mode=figures,round-precision=5]{0.9996446340}&	\num[round-mode=figures,round-precision=5]{0.9996600374}&	\num[round-mode=figures,round-precision=5]{0.9996626181}\\
&	$k^{\sigma}_{4}$&	\num[round-mode=figures,round-precision=5]{0.9993558776}&	\num[round-mode=figures,round-precision=5]{0.9996433030}&	\num[round-mode=figures,round-precision=5]{0.9996588195}&	\num[round-mode=figures,round-precision=5]{0.9996614187}\\
&	$k^{\sigma}_{6}$&	\num[round-mode=figures,round-precision=5]{1.0000000000}&	\num[round-mode=figures,round-precision=5]{0.9998199172}&	\num[round-mode=figures,round-precision=5]{0.9997417355}&	\num[round-mode=figures,round-precision=5]{0.9997437212}\\
&	$k^{\sigma}_{8}$&	\num[round-mode=figures,round-precision=5]{0.9998331665}&	\num[round-mode=figures,round-precision=5]{0.9998179501}&	\num[round-mode=figures,round-precision=5]{0.9998259356}&	\num[round-mode=figures,round-precision=5]{0.9998272884}\\
&	$k^{\sigma}_{10}$&	\num[round-mode=figures,round-precision=5]{0.9993252362}&	\num[round-mode=figures,round-precision=5]{0.9993564994}&	\num[round-mode=figures,round-precision=5]{0.9993849938}&	\num[round-mode=figures,round-precision=5]{0.9993898187}\\
    \hline\hline
    \end{tabular}
    \vspace{0.2cm}
    \caption{Bound success under $J_2$ dynamics assumption (Doppler radar, raw data).}
    \label{tab:Results_bd_success_j2_5s_dr_raw}
\end{table*}

\Crefrange{tab:Results_err_kep_5s_dr_raw}{tab:Results_bd_success_j2_5s_dr_raw} show the results of the application of the proposed method to raw measurements provided by the described Doppler-only bistatic radar. \Cref{tab:Results_err_kep_5s_dr_raw,tab:Results_err_j2_5s_dr_raw} show the trend of the solution accuracy $\varepsilon_{\vb*{x}}$ as a function of noise level and arc length. The same considerations made for optical measurements hold also in this case. Differences with respect to the optical case can be observed when analyzing the trend of the $f_{\vb*{x}}$ parameter. As can be seen, the accuracy on the bounds is always very high, above 99$\%$, regardless of the noise level, arc length, or considered dynamics. This trend can be explained by considering the accuracy of the angular measurements provided by the radar. Compared with optical measurements, there are in fact differences of almost three orders of magnitude between them. Consequently, mismatches in the dynamical model are absorbed by the large bounds estimates that result from large measurement errors, thus ensuring that the true state is almost always included in the provided bounds.

\section{Conclusions}

In this study, three novel \glsentryfull{iod} algorithms were developed for optical, range radar and Doppler-only radar sensors. These methods combine \glsentryfull{da} and \glsentryfull{ads} to obtain the Taylor expansion of the orbital solution as a function of uncertainties in the processed measurements. The algorithms build on previous work presented by the authors based on Keplerian dynamics and extend it to arbitrary dynamical models, thus relaxing any applicability constraint related to the observed arc length and orbital regime. These methods were tested on simulated data obtained by targeting a subset of the NORAD \glsentryfull{leo} population considering an analytical formulation of the $J_2$-perturbed dynamics. A comparison with their Keplerian counterparts shows that the proposed approaches provide more accurate results in terms of both nominal solution and size of the estimated bounds. Future developments include testing the proposed algorithms with other dynamical models and their application to different orbital regimes.

\subsection*{Acknowledgements}

This work is co-funded by the \glsentryfull{cnes} through A. Foss\`a PhD program, and made use of the \glsentryshort{cnes} orbital propagation tools, including the \glsentryshort{pace} library.

\printbibliography

\end{document}